%% file: PALFA-MSP-Timing-v2.tex
\documentclass[paper=portrait,twocolumn,trackchanges]{aastex6}
\usepackage{multirow}
\usepackage{enumitem}
\usepackage{changepage}
\usepackage{ragged2e}
\usepackage{graphicx}
\usepackage{setspace}
\usepackage{amsmath}
\usepackage{threeparttablex}
\usepackage{longtable}
\usepackage{filecontents,catchfile}
\usepackage{float}
\hypersetup{linkcolor=blue,citecolor=blue,filecolor=blue,urlcolor=blue}
\usepackage{url}
\def\gapp{\ifmmode\stackrel{>}{_{\sim}}\else$\stackrel{>}{_{\sim}}$\fi}

\begin{document}
\title{Eight Millisecond Pulsars Discovered in the Arecibo PALFA Survey}

\author{
E.~Parent\altaffilmark{1},
V.~M.~Kaspi\altaffilmark{1},
S.~M.~Ransom\altaffilmark{2},
P. C. C.~Freire\altaffilmark{3},
%B.~Allen\altaffilmark{11,12,22},
%S.~Bogdanov\altaffilmark{13}, 
A.~Brazier\altaffilmark{4},
F.~Camilo\altaffilmark{5}, 
S.~Chatterjee\altaffilmark{4}, 
J. M.~Cordes\altaffilmark{4},
F.~Crawford\altaffilmark{6}, 
J. S.~Deneva\altaffilmark{7}, 
R. D.~Ferdman\altaffilmark{8}, 
J. W. T.~Hessels\altaffilmark{9,10},
%F. A.~Jenet\altaffilmark{21},
%B.~Knispel\altaffilmark{11,12}, 
J.~van Leeuwen\altaffilmark{9,10}, 
A. G.~Lyne\altaffilmark{11}, 
%R.~Lynch\altaffilmark{2}, 
E.~C.~Madsen\altaffilmark{1},
M.~A.~McLaughlin\altaffilmark{12,13},
C.~Patel\altaffilmark{1},
%Z.~Pleunis\altaffilmark{1},
P.~Scholz\altaffilmark{14},
%A.~Seymour\altaffilmark{10}, 
%X.~Siemens\altaffilmark{22}, 
I. H.~Stairs\altaffilmark{15},
%K.~Stovall\altaffilmark{25},
B.~W.~Stappers\altaffilmark{11},
%J.~Swiggum\altaffilmark{22}, 
W.W.~Zhu\altaffilmark{16}
}

\altaffiltext{1}{Dept.~of Physics and McGill Space Institute, McGill Univ., Montreal, QC H3A 2T8, Canada; \url{parente@physics.mcgill.ca}}
\altaffiltext{2}{NRAO, 520 Edgemont Rd., Charlottesville, VA 22903, USA}
\altaffiltext{3}{Max-Planck-Institut f\"{u}r Radioastronomie, Auf dem H\"{u}gel 69, Bonn 53121, Germany}
\altaffiltext{4}{Cornell Center for Astrophysics and Planetary Science, Ithaca, NY 14853, USA}
\altaffiltext{5}{South African Radio Astronomy Observatory, Observatory, 7925, South Africa}
\altaffiltext{6}{Department of Physics and Astronomy, Franklin and Marshall College, Lancaster, PA 17604-3003, USA}
\altaffiltext{7}{George Mason University, resident at the Naval Research Laboratory, Washington, DC 20375}
\altaffiltext{8}{School of Chemistry, University of East Anglia, Norwich Research Park, Norwich NR4 7TJ, UK}
\altaffiltext{9}{ASTRON, the Netherlands Institute for Radio Astronomy, Oude Hoogeveensedijk 4, 7991 PD Dwingeloo, The Netherlands}
\altaffiltext{10}{Anton Pannekoek Institute for Astronomy, University of Amsterdam, Postbus 94249, 1090 GE Amsterdam, The Netherlands}
\altaffiltext{11}{JBCA, School of Physics $\&$ Astronomy, University of Manchester, Manchester, M13 9PL, United Kingdom}
\altaffiltext{12}{Department of Physics and Astronomy, West Virginia University, Morgantown, WV 26501}
\altaffiltext{13}{Center for Gravitational Waves and Cosmology, West Virginia University, Chestnut Ridge Research Building, Morgantown, WV 26505}
\altaffiltext{14}{Dominion Radio Astrophysical Observatory, Herzberg Astronomy $\&$ Astrophysics Research Centre, National Reseach Council Canada, P.O. Box 248, Penticton, V2A 6J9, Canada}
\altaffiltext{15}{Dept. of Physics and Astronomy, University of British Columbia, 6224 Agricultural Road, Vancouver, BC V6T 1Z1 Canada}
\altaffiltext{16}{CAS Key Laboratory of FAST, NAOC, Chinese Academy of Sciences, Beijing 100101, China}

\graphicspath{{figures/}}
\begin{abstract}
We report on eight millisecond pulsars (MSPs) in binary systems discovered with the Arecibo PALFA survey. Phase-coherent timing solutions derived from 2.5 to 5 years of observations carried out at Arecibo and Jodrell Bank observatories are provided. PSR J1921+1929 is a 2.65-ms pulsar in a 39.6-day orbit for which we detect $\gamma$-ray pulsations in archival \textit{Fermi} data. PSR J1928+1245 is a very low-mass-function system with an orbital period of 3.3\,hours that belongs to the non-eclipsing black widow population. We also present PSR J1932+1756, the longest-orbital-period (41.5\,days) intermediate-mass binary pulsar known to date. In light of the numerous discoveries of binary MSPs over the past years, we characterize the Galactic distribution of known MSP binaries in terms of binary class. Our results support and strengthen previous claims that the scatter in the Galactic scale height distribution correlates inversely with the binary mass function. We provide evidence of observational biases against detecting the most recycled pulsars near the Galactic plane, which overestimates the scale height of lighter systems. A possible bimodality in the mass function of MSPs with massive white dwarfs is also reported.    \\
\end{abstract}
\section{Introduction}
Millisecond pulsars (MSPs) are short-period ($P\,\lesssim\,100$ ms) neutron stars that differ from normal pulsars primarily because of their remarkably small spin-down rates ($\dot{P}\,\lesssim\,10^{-17}$) and their different evolutionary histories. Only $\sim\,1\%$ of all normal pulsars are in binary systems, while we observe orbiting companions (predominantly white dwarfs) around $80\%$ of MSPs. Their properties are consistent with them being old pulsars that have been spun up by the accretion of material from a companion. For this reason, they are often referred to as ``recycled pulsars". While the generally accepted definition for MSPs is pulsars having $P\,\lesssim\,30$ ms, we shall apply throughout this work the broader definition proposed in \cite{m17}: pulsars with $P\,<\,100\,$ms whose $\dot{P}$ is smaller than $10^{-17}$. This allows us to include many pulsars having had short-lived recycling phases (e.g., PSR J1753$-$2240) while excluding young pulsars (e.g., the Crab pulsar B0531+21). To date, more than 300 MSPs have been discovered, representing $\sim\,10\%$ of the total known pulsar population \citep{mht+05}, and roughly half of those are found in globular clusters. 
%280 MSPs\footnote{Based on the ATNF Pulsar Database ( \citealt{mht+05}, version 1.59, January 2019) and including the 8 new discoveries presented in this work.}
%Since the discovery of the first MSP (PSR B1937+21) in \citeyear{bkh+82} by \citeauthor{bkh+82},

The extraordinarily stable rotation of most MSPs allows us to use them as celestial clocks in a variety of applications. One notable example is their use as probes of binary motion, enabling precise neutron-star mass measurements and tests of relativistic gravitational theories. Such studies led to the first confirmation of the existence of gravitational radiation (\citealt{tfm79}) predicted by the theory of general relativity. 

The continued discoveries of MSPs is also motivated by the effort to detect nanohertz gravitational waves (GW) emitted by a cosmological population of supermassive black-hole binaries via a ``Pulsar Timing Array'' (PTA, \citealt{hd83,fb90,jb03,haa+10}). This effort relies on the stability of the MSPs forming the array and the expected disturbances in the pulse times-of-arrival by passing gravitational waves. Correlations in those disturbances as a function of angular separation are then inspected to identify potential signals in the form of a stochastic background. Improving PTA sensitivities is most effectively achieved by increasing the number of MSPs with high timing precision \citep{sejr13}. The GW signals detectable by PTAs are in the nanohertz frequency range, many orders of magnitude below the band where laser-interferometer systems, such as aLIGO \citep{aaa+09}, are sensitive. PTA efforts are therefore complementary to interferometric ground-based detectors and can ultimately reveal information about the kinematics, morphology, content and feedback mechanisms of galaxies.

The PALFA pulsar survey uses the 7-beam Arecibo L-Band Feed Array (ALFA) on the Arecibo Observatory (AO) William E. Gordon 305-m telescope in Puerto Rico. It has been surveying low-latitude regions of the Galactic plane at 1.4\,GHz since 2004 \citep{cfl+06,lbh+15}. The great sensitivity of PALFA and its observational parameters (see Section \ref{observations}) are excellent for finding faint, highly dispersed pulsars. To date, the survey has discovered 190 pulsars\footnote{Pulsar discoveries are available on PALFA's discoveries page: \url{http://www.naic.edu/~palfa/newpulsars/}}, including three double neutron star (DNS) systems \citep{lsf+06, vks+15, lfa+16, sfc+18} and 38 highly-recycled MSPs\footnote{Excluding those in this paper, seven PALFA-discovered MSPs have not yet been published, but discovery plots are available on PALFA's discoveries page.} (e.g. \citealt{crl+08,dfc+12,csl+12,kls+15,skl+15,sab+16}) in the Galactic plane. Two PALFA discoveries, PSR J1903+0327 \citep{crl+08} and PSR J0557+1551 \citep{skl+15}, have been included in the NANOGrav PTA \citep{abb+18}. 

In this work, we present timing solutions for eight MSPs in binary systems recently discovered in the PALFA survey: PSRs J1906+0454, J1913+0618, J1921+1929, J1928+1245, J1930+2441, J1932+1756, J1937+1658 and  J2010+3051. In Section \ref{observations}, we describe the discovery and follow-up observations, as well as the timing analysis. The properties of individual pulsars are discussed in Section \ref{properties}, and we examine the Galactic distribution of different MSP-binary classes in Section \ref{Galactic distribution}. Our results are summarized in Section \ref{summary}.
\section{Observations and analysis} \label{observations}
PALFA conducts a survey for pulsars and transients in the Galactic plane ($\lvert b \lvert \,<\,5^{\circ}$). The Mock spectrometers are used as a back-end for the ALFA receiver, providing 322.6\,MHz of bandwidth divided into 960 channels centered at 1375.5\,MHz. The channels are sampled every 64\,$\mu$s. PALFA surveys the two regions of the Galactic plane that lie in the region of the sky between declinations of 0 to 36$^{\circ}$ that can be observed with the telescope: the ``inner'' ($32^{\circ}\,<\,l\,<\,77^{\circ}$) and ``outer'' ($168^{\circ}\,<\,l\,<\,214^{\circ}$) Galaxy regions. The integration times for those longitude ranges are 268 and 180\,sec, respectively.

The survey data are processed by two independent pipelines. The first is a reduced-resolution ``Quicklook'' pipeline \citep{s13} performed in near real-time on site and which enables rapid discovery of bright pulsars. The second is a full-resolution \texttt{PRESTO}-based pipeline \citep{r01}. It processes the data at a Compute Canada/Calcul Qu\'ebec facility hosted at McGill University and searches for pulsars in the Fourier domain. It also searches in the time domain, for pulsars with a Fast-Folding Algorithm \citep{pkr+18}, and has a single-pulse pipeline \citep{pab+18} that searches for Rotating Radio Transients and Fast Radio Bursts. For a more detailed description of the survey, see \cite{lbh+15}.
\begin{figure*}
  \centering
    \includegraphics[width=1.01\textwidth]{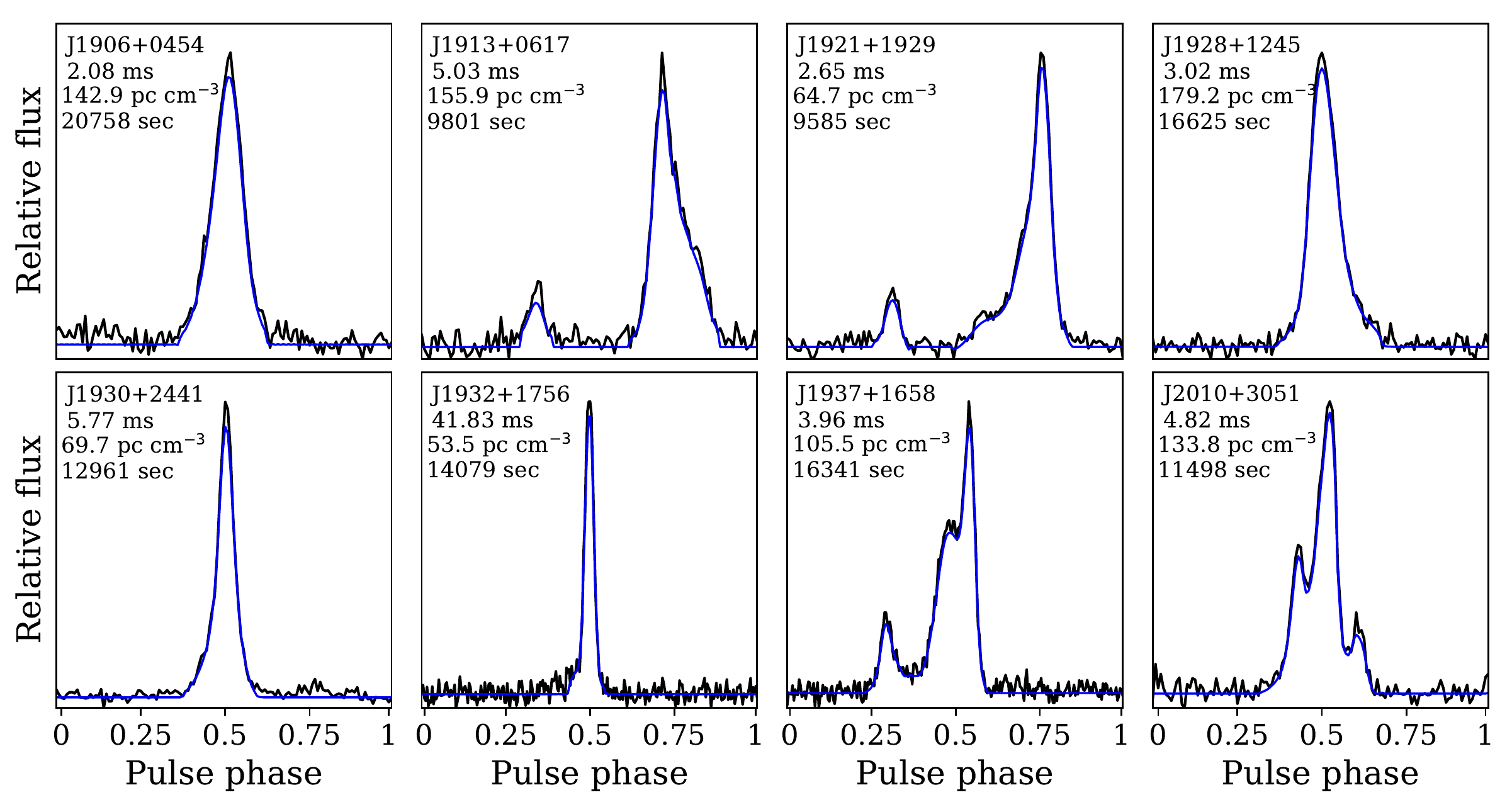}
    \caption{Pulse profiles at 1.4\,GHz of the eight pulsars presented in this work. Black curves correspond to the phase-aligned and summed observations, and the blue curves are the smoothed standard templates used to generate TOAs. One full period is shown in either 128 or 256 phase bins. The name, rotation period, dispersion measure and the total integration time used to generate the profiles are indicated for each pulsar in their respective profile subplot.}
    \label{fig:profs}
\end{figure*}
\subsection{Discovery}
The eight MSPs presented in this work were all discovered in the 268-sec inner-Galaxy pointings between 2014 June and 2016 November. Four of them (PSRs J1921+1929, J1930+2441, J1937+1658 and J2010+3051) were first identified by the ``Quicklook'' pipeline, while PSRs J1906+0454, J1913+0618, J1928+1245 and J1932+1756 were detected by the full-resolution pipeline only. Additional observations were then carried out to confirm the new sources. 
\subsection{Timing observations}
To determine the rotational, astrometric and binary parameters of the systems, follow-up observations for all eight pulsars were conducted at AO using the dual-linear-feed $L$-wide receiver. Three of them (PSRs J1930+2441, J1932+1756 and J2010+3051) had additional data taken with the Lovell Telescope at Jodrell Bank Observatory (JBO). 

Observations at AO have an observing frequency range of 980--1780\,MHz and an average system equivalent flux density of 2.9\,Jy. However, frequencies at the edges of the band are removed and following RFI excision, only $\sim 80\%$ of the data is usable, resulting in an effective band of approximately 1050--1700\,MHz. The Puerto Rico Ultimate Pulsar Processing Instrument (PUPPI) backend was used to record data. Depending on the discovery detection significance and the available time, we initially observed each pulsar for 300, 600 or 900\,sec per session in PUPPI's incoherent-dedispersion search mode. This mode provides total intensity data with 2048 channels read out every 40.96\,$\mu$s. We then downsampled the data to 128 channels, correcting for dispersion delay due to plasma in the interstellar medium before summing channels. Data that were more affected by interference instead had frequency resolution reduced to 256 channels in order to optimize RFI-mitigation. Once initial estimates of the ephemerides were obtained, the PUPPI observations were taken in coherent-dedispersion fold mode. In this mode, full-Stokes polarization data are folded at the predicted pulse period and recorded into 2048 profile bins in real time with 512 frequency channels. The data within each frequency channel are coherently dedispersed to a fiducial best-estimate value. Due to technical issues affecting the local oscillator at Arecibo starting in Summer 2018, some of the most recent data were taken in incoherent search mode.

Observations carried out with the 76-m Lovell Telescope at JBO used a dual-polarization cryogenic receiver having a system noise equivalent flux density on cold sky of 25\,Jy.  A passband from 1350 to 1700\,MHz was processed using a digital filterbank which split the two polarization bands into 0.5-MHz bandwidth channels. The power from each channel was then folded at the nominal topocentric period and the resultant profiles were dedispersed at the dispersion measure of the pulsar and then summed. Each observation was typically of 30--60\,min duration.
\subsection{Timing analysis}
Incoherent search-mode data were cleaned of RFI, dedispersed into barycentric time series and folded using \texttt{PRESTO} tools. From the search mode data taken during the first few months of follow-up observations, we obtained the best-fit period and period derivative of each barycentric fold and applied a model of orbital Doppler shifts with \texttt{PRESTO}'s \texttt{fitorbit} tool. The resulting binary parameters were used as starting points to produce ephemerides. 

The coherently dedispersed data were calibrated, cleaned of RFI and downsampled both in time and frequency, and pulse times-of-arrival (TOAs) were extracted. Profile templates were initially created for each pulsar by fitting Gaussian components to the folded profile of the strongest detection. These standard templates were then used to extract TOAs with \texttt{PSRCHIVE}.
\begin{figure}[t!]
  \centering
    \includegraphics[width=0.51\textwidth,height=19cm]{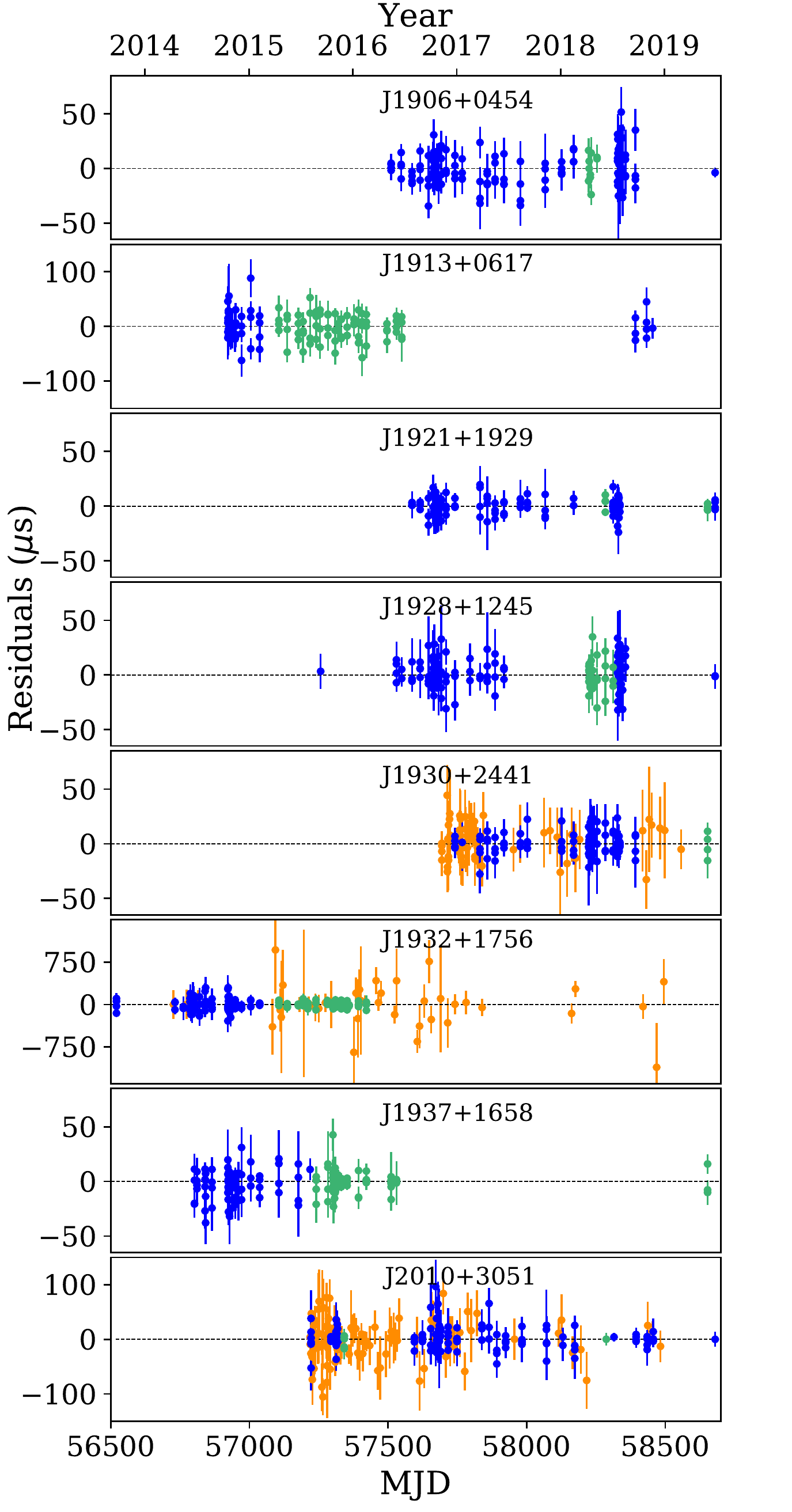}
    \caption{Timing residuals of the best-fit solutions for the eight MSPs presented in this work. The blue and green data points represent observations conducted at AO in search and fold mode, respectively, and the orange points are those obtained with the Lovell Telescope at JBO.}
    \label{fig:res}
\end{figure}
Initial phase-coherent ephemerides for five pulsars (PSRs J1906+0454, J1921+1929, J1928+1245, J1930+2441 and J1932+1756) were obtained using Dracula\footnote{\url{https://github.com/pfreire163/Dracula}} \citep{fr18}, a new algorithm for determining the correct global rotational count with a minimal number of \texttt{TEMPO} iterations for pulsars with sparse timing data. This allowed us to eliminate ambiguity in the number of pulses between observations in the early stages of the phase connection process (i.e., data not sampled densely enough and/or not sampling the orbital phase well enough).

Improved, high-signal-to-noise standard templates were produced for each pulsar by summing frequency- and time-scrunched profiles from multiple observations conducted at AO. Signal-to-noise-based weightings were applied while summing the profiles. Profile baselines were set to zero and the final templates (shown in Figure \ref{fig:profs}) were smoothed using the \texttt{PSRCHIVE} tool \texttt{psrsmooth}, which applies a wavelet smoothing algorithm to create a “noise-free” template profile \citep{cd95,dfg+13}. High-precision TOAs were extracted with \texttt{PSRCHIVE}'s \texttt{pat} tool by fitting for a linear phase gradient in the Fourier domain to determine the shifts between the profiles and the standard template \citep{t92}. 

TOAs extracted from JBO data were obtained using a different profile template than those produced from Arecibo data. We therefore fitted for an arbitrary time offset between the two data sets. 

Formal TOA uncertainties are often underestimated during the extraction procedures, and it is common practice in pulsar timing to increase their uncertainties by some scaling factor (EFAC) to produce a more conservative estimate of errors on the timing parameters. We list these factors in Tables \ref{tab:longtable1} and \ref{tab:longtable2} along with the final \texttt{TEMPO} fits.  

These timing solutions were produced by reducing further the frequency resolution to one to four channels and by folding the data with \texttt{fold\_psrfits}, a component of the \texttt{psrfits\_utils} library\footnote{\url{https://github.com/scottransom/psrfits_utils}} for processing PSRFITS pulsar data files. The DE421 solar-system ephemeris and the UTC(NIST) terrestrial time standard were used in all timing solutions. Figure \ref{fig:res} shows our solution residuals. 

Two binary orbital models were used throughout the analysis: the DD \citep{dd86} and ELL1 \citep{lcw+01}  models.

The DD model consists of an analytic solution for the equation of motion of binary pulsars using the first post-Newtonian approximation of general relativity. It is a theory-independent model that considers effects such as Shapiro delay and aberration due to the pulsar motion. This binary model was applied to two of the eight MSPs presented in this work (PSRs J1913+0618 and J1932+1756), but our current timing precision is insufficient to allow us to measure any post-Newtonian parameters. 

The ELL1 timing model is a modification of the model above, adapted for small-eccentricity binary pulsars where the longitude of periastron, $\omega$, and the time of periastron, $T_0$, are highly correlated. The ELL1 model avoids the covariance between $\omega$ and $T_0$ by parameterizing the orbit with the time of ascending node $T_{asc}$ = $T_0$ - $\omega P_b/ 2 \pi$ and the first and second Laplace parameters, $\epsilon_1 = e$ sin$\omega$ and $\epsilon_2 = e$ cos$\omega$, where $e$ is the eccentricity. This model accounts only for first-order corrections in $e$. It therefore applies to nearly circular orbits where $x e^2$ is much smaller than the error in TOA measurements, where $x$ is the projected semi-major axis. The timing solutions of the six remaining MSPs were obtained using this model.

\subsection{Polarization and flux density measurements} \label{calib}
Polarization calibrations were performed using observations of a noise diode before observing the pulsars in coherent mode with AO's L-wide receiver. For flux calibrations, we used NANOGrav observations of the bright quasars J1413+1509 and J1445+099 \citep{nanograv15}. Both calibrations were conducted with \texttt{PSRCHIVE} tool \texttt{pac} using the \textit{SingleAxis} model, assuming that the polarization of the two receptors are perfectly orthogonal. Polarization profiles for the eight pulsars are shown in Figure \ref{fig:polprof}.

Following data calibration and RFI excision, we combined the data to increase the total linearly polarized flux and searched for rotation measures (RMs) ranging from $-$1000 to 1000\,rad\,m$^{-2}$. Due to the low brightness of the pulsars and the limited number of observations taken in fold mode that could be combined, we could only detect a RM at a significant level for three sources (PSRs J1921+1929, J1928+1245 and J2010+3051). RM and average flux measurements are reported in Tables \ref{tab:longtable1} and \ref{tab:longtable2}. 

Despite showing significant linearly polarized flux and position angle detections for multiple profile bins, the RM value determined from our current dataset for PSR J1937+1658 is consistent with being zero (see Figure \ref{fig:polprof}). Using polarization information of NANOGrav pulsar data taken at AO, \cite{gmd+18} demonstrated that there is an asymmetry in the line-of-sight component of the Galactic magnetic field about a Galactic latitude of 0$^\circ$, producing RM values near zero for the lowest-latitude sources (see Figure 20 in \citealt{gmd+18}). It is therefore likely that PSR J1937+1658 is located in an environment where the Galactic magnetic field has a small component towards our line of sight, resulting in a RM close to zero.

\begin{figure}[h!]
  \centering
    \includegraphics[width=0.52\textwidth]{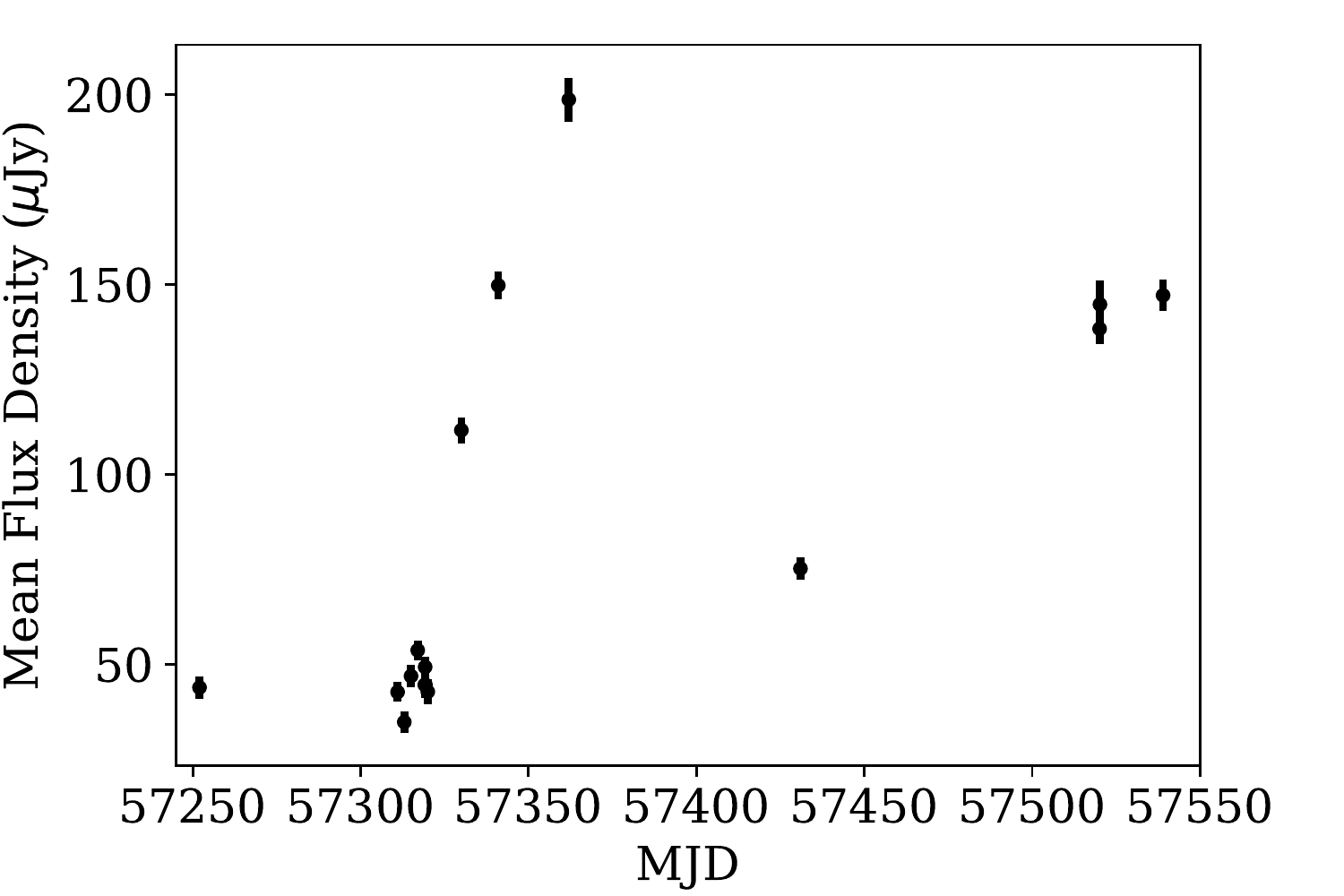}
    \caption{Variation in flux density measurements for PSR J1937+1658, calculated as described in Section \ref{calib}. }
    \label{fig:flux-1937}
\end{figure}

\begin{figure*}[t!]
  \centering
    \includegraphics[width=0.85\textwidth,height=21cm]{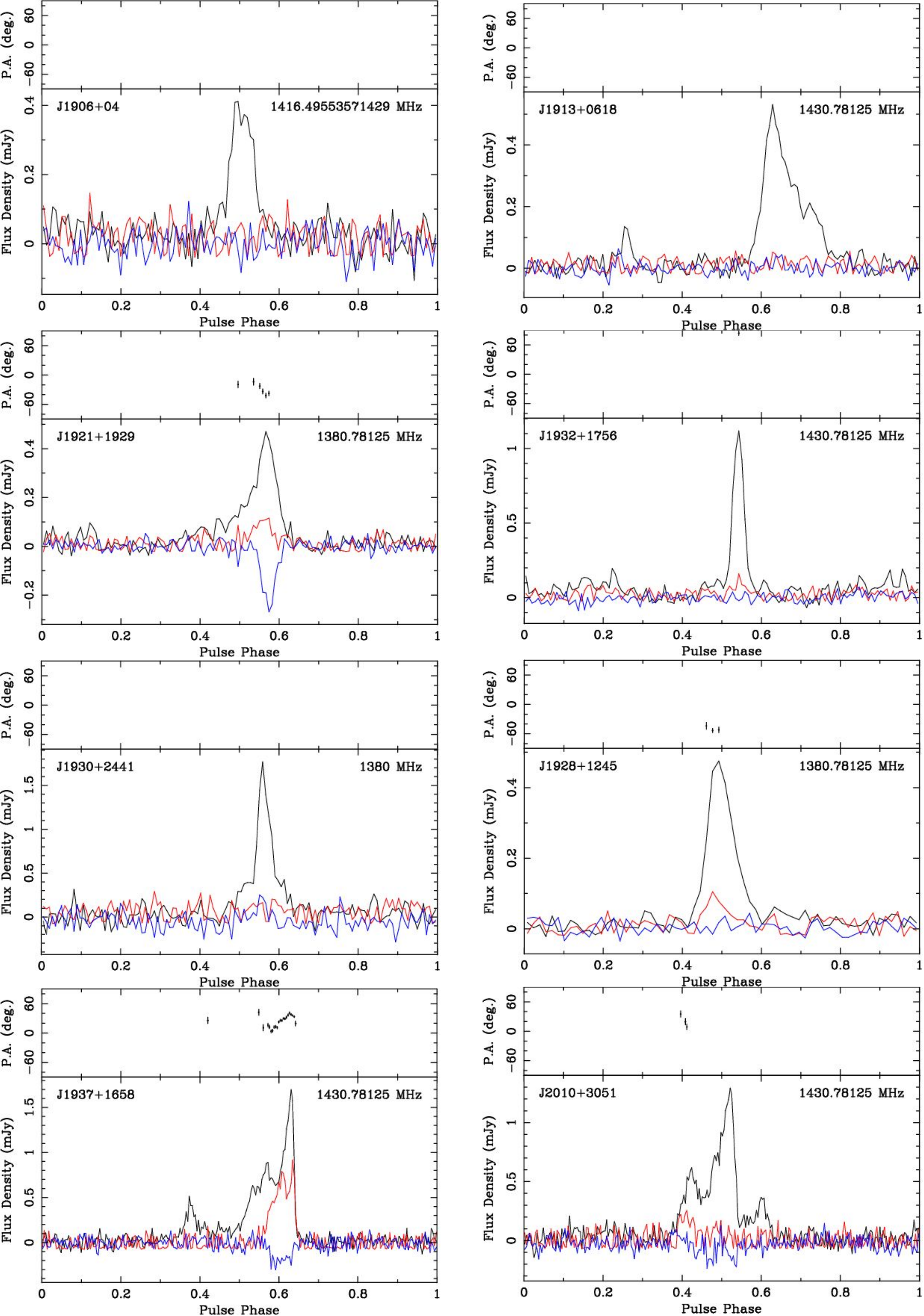}
    \caption{Polarization profiles for the eight MSPs presented in this work. Total (black), linearly polarized (red) and circularly polarized (blue) fluxes are shown in the bottom panel of each plot. The position angle of linear polarization for profile bins with a linear polarization greater than a signal-to-noise ratio of three are displayed in the panel above each profile plot. RM measurements could only be determined marginally for PSRs J1921+1929, J1928+1245 and J2010+3051 (see Tables \ref{tab:longtable1} and \ref{tab:longtable2}). Data for these three pulsars were corrected for Faraday rotation before folding. Only fully-calibrated fold mode data have been folded to produce these profiles. Note that while the profile PSR J1937+1658 shows significant linearly polarized flux, the RM determined from our current dataset is consistent with being zero. Larger datasets will allow us to obtain more precise RM measurements. }
    \label{fig:polprof}
\end{figure*}

We note a flux enhancement for PSR J1937+1658 during the month of November 2015, before which the flux was fairly stable around 45\,$\mu$Jy (see Figure \ref{fig:flux-1937}). It then reached $\sim$\,200\,$\mu$Jy by the end of the month, and the flux subsequently remained at $\sim$\,140\,$\mu$Jy. 
%% ---------   TABLES - TIMING SOLUTIONS  ----------------
\clearpage
\setlength{\tabcolsep}{0.75mm}
\LTcapwidth=2.1\linewidth
\newcommand{\longtableone}{\input{table-timing-1.tex}}% table.tex > \mytable
\begin{spacing}{.85}
\begin{ThreePartTable}
\begin{TableNotes}
\item [a] Observed period derivative obtained from the timing analysis.
\item [b] Calculated using the derived intrinsic period derivative.
\item [c] DM-derived distance using the YMW16 Galactic electron density model.
\item [d] Pulse width at 50$\%$ of peak at 1400\,MHz.
\item [e] Calculated at 1400\,MHz.
\item [f] Daily averaged and weighted rms residuals.
\item [g] Scaling factor used to increase TOA uncertainties.
\end{TableNotes}
\begin{longtable*}{p{6.5cm} c c c c} 
    \caption{\normalsize Timing parameters of PALFA-discovered MSPs. Numbers in parentheses are \texttt{TEMPO}-reported uncertainties in the last digit quoted.}
    \label{tab:longtable1} \\ \hline
    \input{table-timing-1.tex}
\end{longtable*} 
\end{ThreePartTable}
\end{spacing}
%---------------------------- TABLE 2
\setlength{\tabcolsep}{0.75mm}
\LTcapwidth=2.1\linewidth
\newcommand{\longtabletwo}{\input{table-timing-2.tex}}% table.tex > \mytable
\begin{spacing}{.85}
\begin{ThreePartTable}
\begin{TableNotes}
\item [a] Observed period derivative obtained from the timing analysis.
\item [b] Calculated using the derived intrinsic period derivative.
\item [c] DM-derived distance using the YMW16 Galactic electron density model.
\item [d] Pulse width at 50$\%$ of the peak at 1400\,MHz (calculated on the primary peak when multiple components are present in the profile).
\item [e] Calculated at 1400\,MHz.
\item [f] Daily averaged and weighted rms residuals.
\item [g] Scaling factor used to increase TOA uncertainties.
\end{TableNotes}
\begin{longtable*}{p{6.5cm} c c c c}
    \caption{\normalsize Timing parameters of PALFA-discovered MSPs. Numbers in parentheses are \texttt{TEMPO}-reported uncertainties in the last digit quoted.} 
    \label{tab:longtable2}  \\ \hline 
    \input{table-timing-2.tex} 
\end{longtable*}
\end{ThreePartTable}
\end{spacing}
%% ---------   END TABLES ----------------
\section{Properties of new discoveries} \label{properties}
\begin{figure}[b!]
  \centering
    \includegraphics[width=0.5\textwidth]{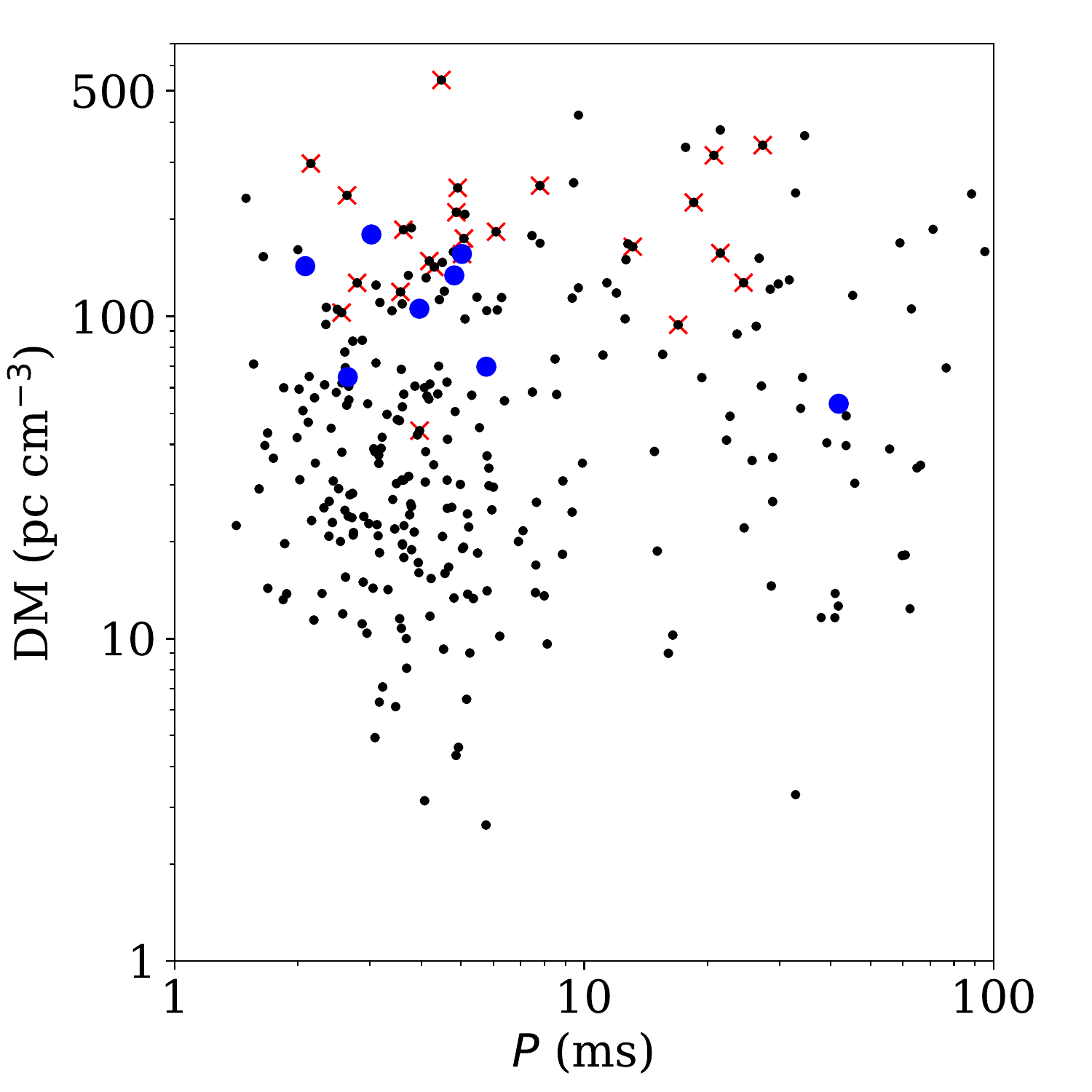}
    \caption{DM values of recycled pulsars (black dots) having known period derivatives, where the PALFA-discovered sources are marked with red ``X" symbols. Blue points are the MSPs presented in this work. This figure illustrates the important contribution of the PALFA survey to the known population of high-DM MSPs, especially at short rotation periods. }
    \label{fig:DMvsP}
\end{figure}

\begin{figure}[b!]
 %\raggedleft
    \includegraphics[width=0.54\textwidth]{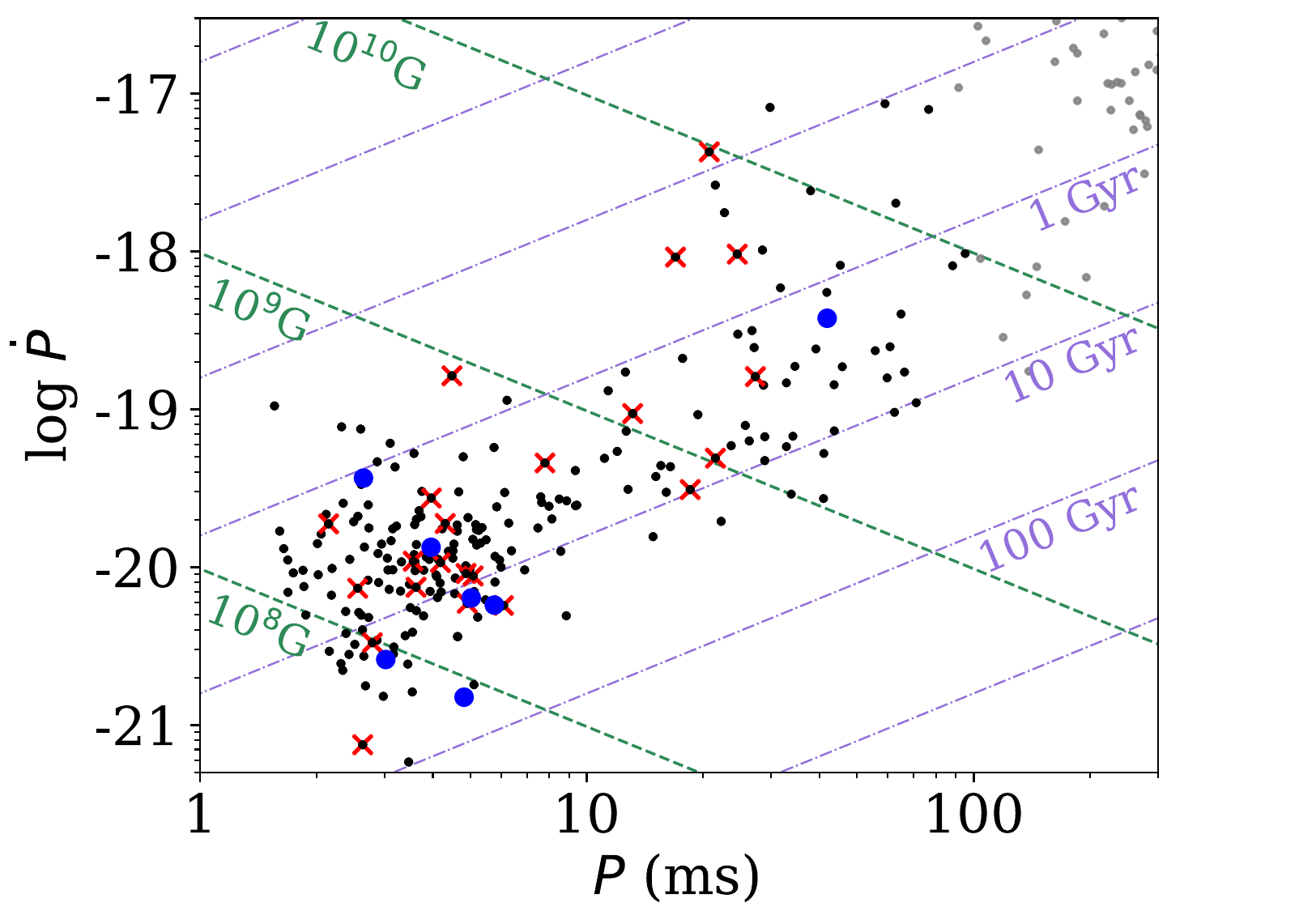}
    \caption{Period-period derivative ($P$-$\dot{P}$) diagram for pulsars with $P\,<\,300$ ms. Black dots are MSPs ($P\,<$\,100\,ms and $\dot{P}\,<\,1\,\times\,10^{-17}$),  while the grey dots are normal (non-recycled) pulsars. Pulsars discovered by the PALFA survey are marked with red ``X" symbols, and the eight MSPs presented in this paper are plotted with blue points. Green dashed lines correspond to derived surface magnetic field strength at constant values of $10^8, 10^9$ and $10^{10}$\,G. Pulsar characteristic ages of 1, 10 and 100\,Gyr are represented by purple dashed lines. Data were taken from the ATNF Catalogue (version 1.60).}
    \label{fig:ppdot}
\end{figure}

All pulsars presented in this paper were found at Galactic longitudes between 39$^{\circ}$ and 68$^{\circ}$, and within 3$^{\circ}$ of the Galactic plane. Except for PSR J1932+1756 ($P$\,=\,41.8 ms), they all have rotation periods shorter than 6\,ms, adding to the low-B field population occupying the lower-left corner of the $P$-$\dot{P}$ diagram (see Figure \ref{fig:ppdot}). Their DM values range from 53 to 179\,pc\,cm$^{-3}$, and five of them have DMs greater than 100\,pc\,cm$^{-3}$ (see Figure \ref{fig:DMvsP}). Distances were estimated using both the NE2001 Galactic density electron model \citep{cl02} and the \citet[][hereafter YMW16]{ymw16} model. The distances estimated with the two models agreed with each other well within the estimated 25\% uncertainty for all pulsars. Throughout the rest of this work, we will refer to distance predictions from the YMW16 model to be consistent with the ATNF Pulsar Catalogue\footnote{ \url{http://www.atnf.csiro.au/research/pulsar/psrcat/}, version 1.60} \citep{mht+05}, which applied that model to all database entries relying on distance estimates (as of 2016 November).    

Measured and derived pulsar parameters are reported in Tables \ref{tab:longtable1} and \ref{tab:longtable2}, and some of their most notable properties are discussed below. We first present the derivations of intrinsic spin-down rates and upper limits on proper motions and distances.

\subsection{Intrinsic period derivative calculations} \label{pdot_int}
Aside from the intrinsic spin-down rate $\dot{P}_{\rm int}$ of the pulsar, there are several effects contributing to the observed period derivative $\dot{P}_{\rm obs}$, and those can be significant for MSPs given their small $\dot{P}_{\rm int}$. In most cases, the kinematic effects contribute the most to the deviation of the $\dot{P}_{\rm obs}$ from $\dot{P}_{\rm int}$. Apparent accelerations arise as a result of transverse motions (Shklovskii effect; \citealt{s70}) and relative accelerations between the pulsar and the Solar System Barycenter \citep{dt91}. Proper motion and distance measurements allow us to calculate the kinematic contributions to the observed period derivative $\dot{P}_{\rm obs}$, and therefore predict $\dot{P}_{\rm int}$:
\begin{equation}
    \dot{P}_{\rm int} =  \dot{P}_{\rm obs} - \dot{P}_{\rm gal} - \dot{P}_{\rm shk}, 
\end{equation}
where $\dot{P}_{\rm gal}$ includes both a vertical component of the Galactic acceleration of the pulsar and the line-of-sight acceleration due to Galactic differential rotation between the Solar System and the pulsar. $\dot{P}_{\rm shk}$ is the result of the Shklovskii effect, and is calculated using the following equation \citep{s70}:
\begin{equation}
    \dot{P}_{\rm shk} = \frac{P}{c} \mu_T^2 D.
\end{equation}
The terms $P$, $\mu_T$ and $D$ are the rotation period, the proper motion and the distance, respectively. 

When using $\mu_T$ directly from the timing analysis and the DM-distances to calculate $\dot{P}_{\rm int}$, we a obtain negative value for one of the pulsars. Using $\dot{P}_{\rm obs}$ as an upper limit for the kinematic terms ($\dot{P}_{\rm gal} + \dot{P}_{\rm shk}$), we can better estimate the uncertainty on our measurements and determine the most likely distance and proper motion for each pulsar.

Similarly to the technique performed by \cite{nbb+14}, the contribution from each term in Equation (1) and their associated uncertainties were estimated via a Monte Carlo rejection sampling. For each pulsar, we generated sampling combinations of $\dot{P}_{\rm obs}$, $\mu_T$ and $D$, each drawn from probability density functions of normal distributions having means equal to the measured values. Uncertainties of 25$\%$ were used to construct the distance distributions. For MSPs where we did not detect significant proper motions, we used the measured upper limits as constraints. $\dot{P}_{\rm gal}$, $\dot{P}_{\rm shk}$ and $\dot{P}_{\rm int}$ were computed for each combination, where we used the approximation from \cite{lwj+09} (valid for Galactic height $<\,1.5$\,kpc) for the vertical component contribution to the Galactic acceleration, and Equation (5) of \cite{nt95} for the line-of-sight relative acceleration. Values used for the Galactocentric distance of the Solar System and the rotational speed of the Galaxy at the Solar System are those derived in \cite{rmb+14}. 

In the rejection process, we assumed that MSPs are not gaining rotational energy (i.e., $\dot{P}_{\rm int}\,>\,0$) and that no additional gravitational potential significantly affects the observed period derivatives. Sampling combinations producing $D\,\leq$\,0 or $\mu_T\,<$\,0 were also rejected. The resulting mean values and their associated errors of the various estimations of the period derivatives (see Table \ref{tab:pdot}) were calculated from 10$^6$ valid combinations of our Monte Carlo iterations. We note that since estimations of Shklovskii accelerations correlate strongly with the measured $\mu_T$ and that most of our timing measurements only weakly constrain this parameter, the low-significance $\dot{P}_{\rm shk}$ values we report are expected. The most likely values produced by the simulations for the proper motions, $\mu_{T,MC}$, are all consistent with measured $\mu_T$ values produced by the timing analysis, and the relative errors are reduced by $\sim\,20\%$. All distance estimates produced by our Monte Carlo simulations are consistent with the values predicted by both the NE2001 and YMW16 models with average relative errors between 25 and 30\%.
\input{table_mu_d_v.tex}
\input{table_pdot.tex}

\subsection{Timing results}
\begin{figure*}
  \centering
    \includegraphics[width=1.01\textwidth]{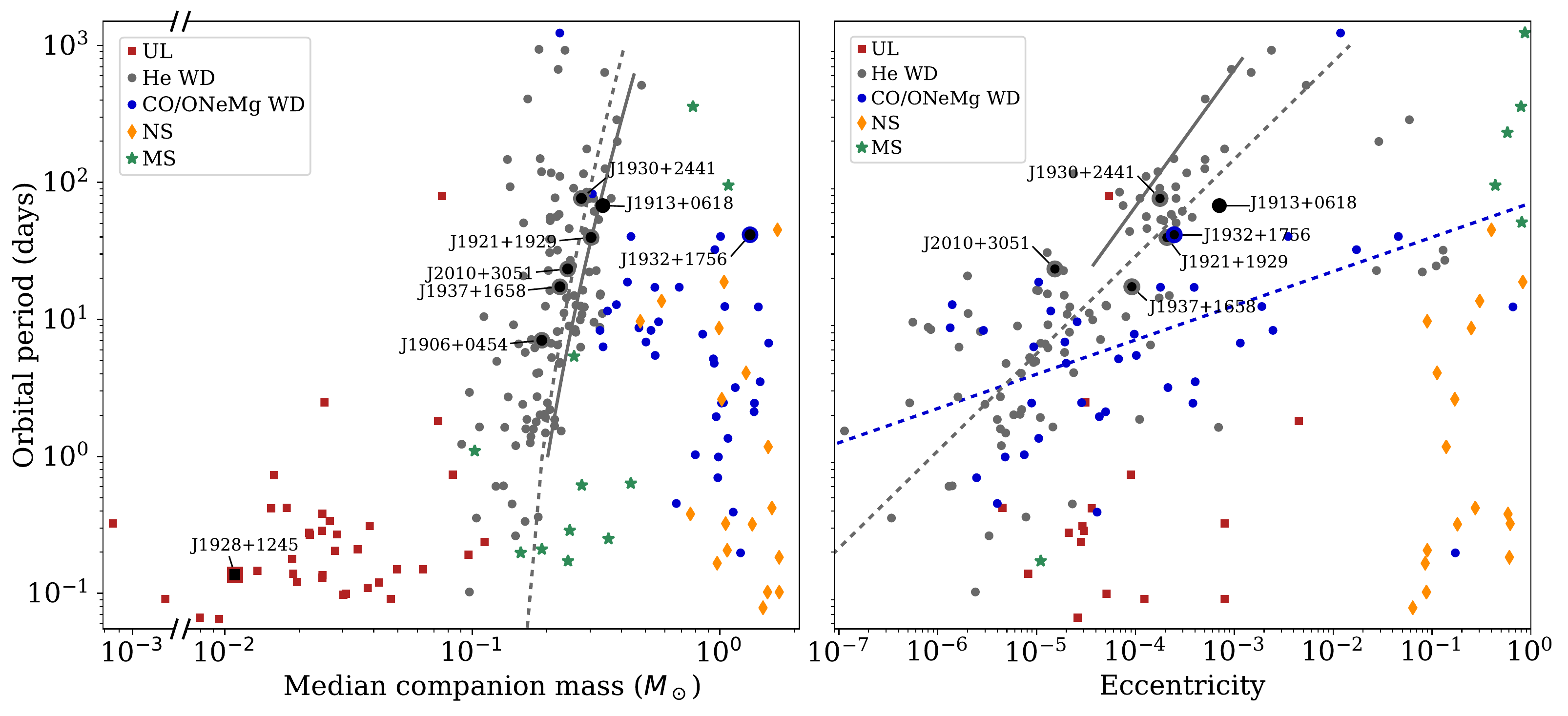}
    \caption{Orbital properties of Galactic MSPs in binary systems, taken from the ATNF Catalogue and the online Galactic MSP Catalog\textsuperscript{a}. The different colors refer to the different companion types (see legend). UL refers to ``Ultra-Light'' companion, ``NS'' to neutron star and ``MS'' to main sequence star. MSPs presented in this work are labelled and represented by black-filled symbols. \textit{Left panel}: Orbital period as a function of derived median companion mass of stellar companions. The solid grey line is the theoretical prediction derived by \cite{ts99} for MSP-He WDs systems formed from LMXBs, and the dashed line is the correlation found by \cite{hwh+18} for the same class. \textit{Right panel}: Eccentricity $e$ versus orbital period $P_b$ for Galactic MSPs in binary systems.
    The prediction for the eccentricity of MSP-He WD systems with a given $P_b$ derived in \cite{p92}, where the residual ellipticity is explained as a ``freeze-out'' during the companion transition from a red giant to a WD, is shown by the grey solid line. Dashed grey and blue lines are the statistical correlations obtained by \cite{hwh+18} for the He and CO/ONeMg WD binary classes, respectively. }
    \begin{flushleft}   \footnotesize\textsuperscript{a}\url{http://astro.phys.wvu.edu/GalacticMSPs/GalacticMSPs.txt}\end{flushleft}
    \label{fig:binaries_params}
\end{figure*}
\subsubsection*{PSR J1906+0454}
With a rotation period of 2.08\,ms, PSR J1906+0454 is the most rapidly rotating MSP of the eight, and it is the second most rapidly rotating object found by the PALFA survey. The pulse width at half maximum (W50) is 0.20\,ms, corresponding to a pulse duty cycle of 10$\%$. According to the YMW16 model, its DM of 142.8\,pc\,cm$^{-3}$ predicts a distance of 4.1\,kpc. 
%Following the procedure described in Section \ref{pdot_int}, we derive a 2$\sigma$ distance upper limit of 6 kpc. 

It is in a nearly circular 7.04-day orbit, the second shortest orbital period of the pulsars presented here. Assuming a pulsar mass $M_{p}$ of 1.40\,$M_\odot$ and an orbital inclination angle $i$ of 60$^{\circ}$, the median mass of the companion is 0.191\,$M_\odot$. The proximity of PSR J1906+0454 to the companion mass-orbital period relation predicted by \cite{ts99} for MSP-WD systems born from low-mass X-ray binaries (LMXBs) suggests that the companion is a low-mass Helium white dwarf (He WD, see Figure \ref{fig:binaries_params}). This hypothesis is further supported by the rapid rotation of the pulsar, indicating a past long-term accretion phase.

A total proper motion detection $\mu_T$ of 15\,$\pm$\,3\,mas\,yr$^{-1}$ was obtained from the timing analysis (see Table \ref{tab:longtable1}), corresponding to a transverse velocity of 300\,$\pm$\,90\,km\,s$^{-1}$ for a 4.1-kpc distance. Our Monte Carlo simulations suggest a somewhat lower value of 199$_{-38}^{+45}$\,km\,s$^{-1}$.

After correcting for apparent accelerations arising from the Shklovskii effect and the Galactic potential, our calculations of the intrinsic period derivative give $\dot{P}_{\rm int} = 0.07_{-0.02}^{+0.04}\,\times\,10^{-20}$. The Shklovski effect is the most significant and accounts for nearly 85\% of the observed spin-down.

Our timing solution spans 3.2\,years and has a post-fit rms timing residual of 12.15\,$\mu$s.

\subsubsection*{PSR J1913+0618}
PSR J1913+0618 has a spin period of 5.03\,ms, an observed spin period derivative of 9.61\,$\times\,10^{-21}$ and a DM of 155.99\,pc\,cm$^{-3}$. The DM-implied distance is 5.9\,kpc. A 9.4$\%$ duty cycle main pulse with a weak interpulse are observed for this pulsar.

The orbital period of this system is 67.7\,days and it has an eccentricity of $7\,\times\,10^{-4}$. Excluding double-neutron-star systems (DNSs), PSR J1913+0618 belongs to the top 15$\%$ most eccentric recycled pulsars in the Galactic field. The median companion mass is 0.34\,$M_\odot$. According to the classification proposed by \cite{tlk12} and the position of this pulsar in both the orbital period-eccentricity and companion mass-orbital period diagrams (Figure \ref{fig:binaries_params}), the nature of the companion is consistent with being either a massive He WD or a low-mass carbon-oxygen (CO) or oxygen-neon-magnesium (ONeMg) WD. 

We measured $\mu_T=9\,\pm$\,2\,mas\,yr$^{-1}$, and the large distance of the pulsar implies a large transverse velocity of 250\,$\pm$\,83\,km\,s$^{-1}$, making PSR J1913+0618 among the top ten highest transverse-velocity
MSPs known in the Galactic plane\footnote{According to the ATNF Catalogue, version 1.60}. The MC sampling suggests similar values (see Table \ref{tab:pm}). 

Our timing solution spans 4.2\,years and has an rms residuals of 19.78\,$\mu$s. The complete set of parameters obtained from our timing analysis is given in Table \ref{tab:longtable1}. 

\subsubsection*{PSR J1921+1929}
PSR J1921+1929 is a 2.45-ms pulsar with a DM of 64.8\,pc\,cm$^{-3}$ that is spatially coincident with a \textit{Fermi} unassociated point source \citep{fermi4yr}. We folded $\gamma$-ray photons with our timing solution (Table \ref{tab:longtable1}) and detect pulsations. Details on our analysis of the high-energy emission are presented in Section \ref{counterparts}. 

After correcting for the Shklovskii effect and accelerations in the Galactic potential, we obtain $\dot{P}_{\rm int}\,=\,3.65_{-0.04}^{+0.03}\,\times \,10^{-20}$. This corresponds to a spin-down power $\dot{E}$ of 8.1\,$\times\,10^{34}$\,erg\,s$^{-1}$, a typical value for radio-emitting $\gamma$-ray MSPs (see e.g. \citealt{fermi13}). 

The binary system has an orbital period of 39.6\,days and a median companion mass of 0.302\,$M_\odot$, in very good agreement with the mass predicted by \cite{ts99} for He WD companions. PSR J1921+1929 has orbital parameters that are consistent with being born from LMXBs, and combined with its short rotation period, the companion is most likely a He WD if we assume an inclination angle of 60$^\circ$.

The measured composite proper motion is 12\,$\pm$\,1\,mas\,yr$^{-1}$, but the current data do not allow us to measure a parallax distance (3$\sigma$ upper limit of 9\,mas). 

We achieved a post-fit rms timing residual of 4.43\,$\mu$s from three years of data (see Table \ref{tab:longtable2}), the smallest residuals of the eight pulsars presented in this work. This source could be of possible interest for PTAs given its relatively low DM and rapid rotation.

\subsubsection*{PSR J1928+1245}
PSR J1928+1245 has a spin period of 3.02\,ms and the largest DM (179.2\,pc\,cm$^{-3}$) of the eight MSPs presented in this paper, corresponding to a predicted distance of 6.1\,kpc. It is one of the ten most dispersed MSPs known. PSR J1928+1245 is in a short 3.28-hour orbit having a projected semi-major axis $x$ of 0.019\,lt-s. We also measure a first derivative of the orbital period, $\dot P_B$, at 34$\,\pm\,5\times 10^{-12}$. Based on those orbital parameters, the implied mass function (i.e., the lower limit on the mass of the unseen companion) is 3.9\,$\times\,10^{-7}\,M_\odot$, one of the lowest among known MSPs. The minimum and median companion masses are 0.009 and 0.011\,$M_\odot$, respectively. Those are properties commonly shared by ``black widow'' pulsars \citep{el88,fbb+90}: close binary systems in which the very low-mass companion (a non-degenerate or partially degenerate stellar core) is ablated by the pulsar wind. Black widows are typically found to have periods of a few ms, in orbits with $P_b\,\lesssim\,1$\,day and have been suggested to be the progenitors of solitary MSPs. One can see by the position of PSR J1928+1245 in the left panel of Figure \ref{fig:binaries_params} (black-filled square in the lower left corner) that it belongs to the black widow population. If the orbital inclination of such systems is high, plasma surrounding the companion may cause eclipses of the radio pulsar as it passes superior conjunction. 

Our observations that cover orbital phases corresponding to superior conjunction show no evidence for eclipsing or flux reduction. The non-detection of eclipses combined with a very low mass function could imply that the system is far from edge-on, as is suggested by the apparent correlation between the orbital inclination and the presence of eclipses \citep{f05,nbb+14,goc+19}.

Current data do not allow us to measure a proper motion for the pulsar, but the best-value $\mu_{T,MC}$ we obtain is 4.4\,$\pm$\,1.3\,mas\,yr$^{-1}$.

Our solution spans 3.9\,years and has a post-fit rms residual of 10.35\,$\mu$s. All timing parameters for PSR J1928+1245 are listed in Table \ref{tab:longtable1}.

\subsubsection*{PSR J1930+2441}
PSR J1930+2441 is a 5.77-ms pulsar and has DM of 69.6\,pc\,cm$^{-3}$, implying a distance of 3.3\,kpc. It is also the most recent discovery (2016 November) we present in this work. The orbital period of this binary system is 76.4\,days and the companion has a derived median mass of 0.28\,$M_\odot$. The orbital parameters and the short spin period suggest that a He WD is the most likely companion type for this pulsar. 

We measure a total proper motion $\mu_T$ of 10\,$\pm$\,3\,mas\,yr$^{-1}$ and a parallax distance of 16\,$\pm$\,7\,mas. Values produced by our Monte Carlo simulations are consistent with timing measurements and the estimated transverse velocity is 155\,$\pm$\,60\,km\,s$^{-1}$.  However, given the short time-span of the dataset available for this pulsar, future proper motion measurements may vary from the quoted values given that they are highly covariant with other parameters. 

Its timing solution (see Table \ref{tab:longtable2}) has a post-fit rms residual of 8.89\,$\mu$s and spans 2.6\,years.

\subsubsection*{PSR J1932+1756}
PSR J1932+1756 has the longest rotation period of the eight MSPs with $P$\,=\,41.83\,ms. It has a DM of 53.2\,pc\,cm$^{-3}$ and an implied distance of 2.1\,kpc.    

The binary system has a 41.5-day period and for a 1.4\,$M_\odot$ MSP, the companion must have a mass of at least 1.1\,$M_\odot$ (median mass of 1.3\,$M_\odot$). The low eccentricity ($e$\,=\,2.5\,$\times\,10^{-4}$) of the orbit suggests that this massive companion is not a neutron star.

Subtracting the contribution of the Galactic acceleration and the Shklovskii effect to the observed $\dot{P}_{\rm obs}$, we obtain a $\dot{P}_{\rm int}\,$ of $\sim  \,3.8\,\times\,10^{-19}$ and infer a surface magnetic field $B$ of 4\,$\times\,10^{9}$\,G. Those parameters indicate that the pulsar is old and has been recycled.  

PSR J1932+1756 shares the characteristics of the intermediate-mass binary pulsar (IMBPs) class \citep{clm+01}, which have spin periods of a few tens of milliseconds, median companion masses $\geq\,0.5\,M_\odot$ and eccentricities larger than the low-mass binary pulsars. This pulsar has the longest orbital period of the known IMBPs. For systems with orbital periods longer than a few days, the most probable formation channel involves a short-lived phase of super-Eddington mass transfer on a timescale of a few Myr. During this process, the system also has to avoid a common-envelope phase \citep{tvs00}, otherwise frictional forces would result in the neutron star spiraling in and collapsing into a black hole. Heavy WDs having shallow convective envelopes can provide a formation channel for such systems when P$_B\,\sim$\,3--50\,days \citep{tvs00}.  

We obtained a timing rms residual of 51.92\,$\mu$s, spanning a 5.4-year period (see Table \ref{tab:longtable2}). If the orbital inclination is favorable, PSR J1932+1756 could be a potential candidate for a future Shapiro delay measurement. However the rather large timing residual would make this measurement difficult. Our timing data currently do not allow us to detect any Shapiro delay.  

\subsubsection*{PSR J1937+1658}
PSR J1937+1658 has a spin period of 3.96\,ms, a DM of 105.8\,pc\,cm$^{-3}$ implying a distance of 3.3\,kpc, and has a complex pulse profile structure. It is a 17.3-day orbit with a derived eccentricity of $e\,=\,9\,\times\,10^{-5}$ and a median companion mass of 0.23\,$M_\odot$. A He WD is the most likely companion for this pulsar, as suggests its position on the orbital period-companion mass diagram in Figure \ref{fig:binaries_params}.

We measure a small proper motion $\mu_T$ of 5\,$\pm$\,1\,mas\,yr$^{-1}$, corresponding to a small transverse velocity of 77\,$\pm$\,24\,km\,s$^{-1}$. Similar values are obtained from the MC rejection technique.

Our timing solution has an rms residual of 6.88\,$\mu$s and spans five years. Timing parameters are listed in Table \ref{tab:longtable2}.   

\subsubsection*{PSR J2010+3051}
PSR J2010+3051 is a 4.8-ms pulsar with a DM of 133.8\,pc\,cm$^{-3}$ and is the most distant (DM-implied distance of 6.5\,kpc) of the MSPs presented in this work. The orbital period of this nearly circular orbit binary is 23.4\,days with a median companion mass of 0.243\,$M_\odot$. This system shares orbital properties similar to those of PSR J1937+1658, and is located near that source on the period-companion mass diagram (Figure \ref{fig:binaries_params}).

Our best-fit proper motion ($\mu_{T,MC}\,=10.1\,\pm$\,0.5\,mas\,yr$^{-1}$) and distance ($D$\,=\,5.1\,kpc) resulting from the Monte Carlo sampling implies a large transverse velocity of 244\,$\pm$\,38\,km\,s$^{-1}$. Correcting for the kinematic contributions to $\dot{P}_{\rm obs}$ results in $\dot{P}_{\rm int}\,\sim\,1.2\,\times\,10^{-21}$. 

We achieve a post-fit rms residual of 13.31\,$\mu$s for a solution spanning a four-year period (see Table \ref{tab:longtable2}). 

\subsection{Multiwavelength counterparts} \label{counterparts}
Archival optical and infrared data\footnote{\url{https://irsa.ipac.caltech.edu/}} were examined to identify possible counterparts to the eight MSPs. Point sources from either the Two Micron All-Sky Survey (2MASS; \citealp{scs+06}) or the GAIA DR1 \citep{gaia} catalogues that have positions consistent with being associated with PSRs J1913+0618, J1928+1245 and J1932+1756 were identified. However, even when adopting the 95$\%$ lower limit for the pulsar distances with minimal visual extinction and the companion being the hottest/most luminous WDs possible, all the possible infrared or optical counterparts have apparent magnitudes too bright to be consistent with being the true pulsar system counterpart. Considering the large distances of the pulsars presented here and their locations within crowded fields where the large amounts of dust and gas cause considerable optical extinction ($A_v$ between $\sim$\,0.75 and 2.5\,mag for the measured hydrogen column densities), finding associations to our systems is unlikely. Given that PSR J1932+1756 is the nearest source and has a relatively small characteristic age ($\tau_c\,\sim$\,1.5\,Gyr), it should be an interesting prospect for optical or infrared counterpart detection from future follow-up observations, however the WD companion is expected to have a mass larger than 1\,$M_\odot$ and therefore should be small in size, making detection unlikely. Even by considering the most optimistic scenario for the WD properties and assuming little extinction, apparent magnitudes in near-infrared would still be $\gapp$\,25.

We have inspected X-ray catalogs from the missions \textit{ASCA} \citep{smk+01}, \textit{Chandra} (CXOGSG,\citealt{wlq+16}), \textit{ROSAT} (2RXS, \citealt{bft+16}),  \textit{RXTE}\footnote{\url{https://heasarc.gsfc.nasa.gov/docs/xte/recipes/mllc_start.html}}, \textit{Swift}\footnote{\url{https://heasarc.gsfc.nasa.gov/W3Browse/swift/swiftmastr.html}} and \textit{XMM-Newton} (XMMSL1, \citealt{sre+08}) for possible counterparts. No source with position within 2$^\prime$ from the MSP locations was found. 

We also searched for high-energy counterparts using the \textit{Fermi} Large Array Telescope (LAT) 4-year Point Source Catalog\footnote{ \url{https://fermi.gsfc.nasa.gov/ssc/data/access/lat/4yr_catalog/}} \citep{fermi4yr}. The 4$\sigma$-significance \textit{Fermi} unassociated source 3FGL J1921.6+1934 has a position that coincides with PSR J1921+1929 (within the \textit{Fermi} position error of 4$^\prime$). This source has a power-law-type spectrum and  $\gamma$-ray spectral index of $-$2.5, properties that are typical for pulsars emitting high-energy photons. Using the DM-implied distance for this pulsar and the energy flux reported in the LAT catalog (11.7\,$\times\,10^{-12}$\,erg\,cm$^{-2}$\,s$^{-1}$ in the 0.1--100\,GeV range), we calculate a $\gamma$-ray luminosity $L_\gamma$ of 8\,$\times\,10^{33}$\,erg\,s$^{-1}$ at energies 0.1--100\,GeV. The corresponding $\gamma$-ray efficiency $\eta\,=\,L_\gamma$/$\dot{E}$ is approximately 0.1, a typical value according to the second catalog of gamma-ray pulsars \citep{fermi13}. We note however that this value depends on the predicted distance to the pulsar, which has large uncertainty.  

\begin{figure}
  \centering
    \includegraphics[width=0.475\textwidth]{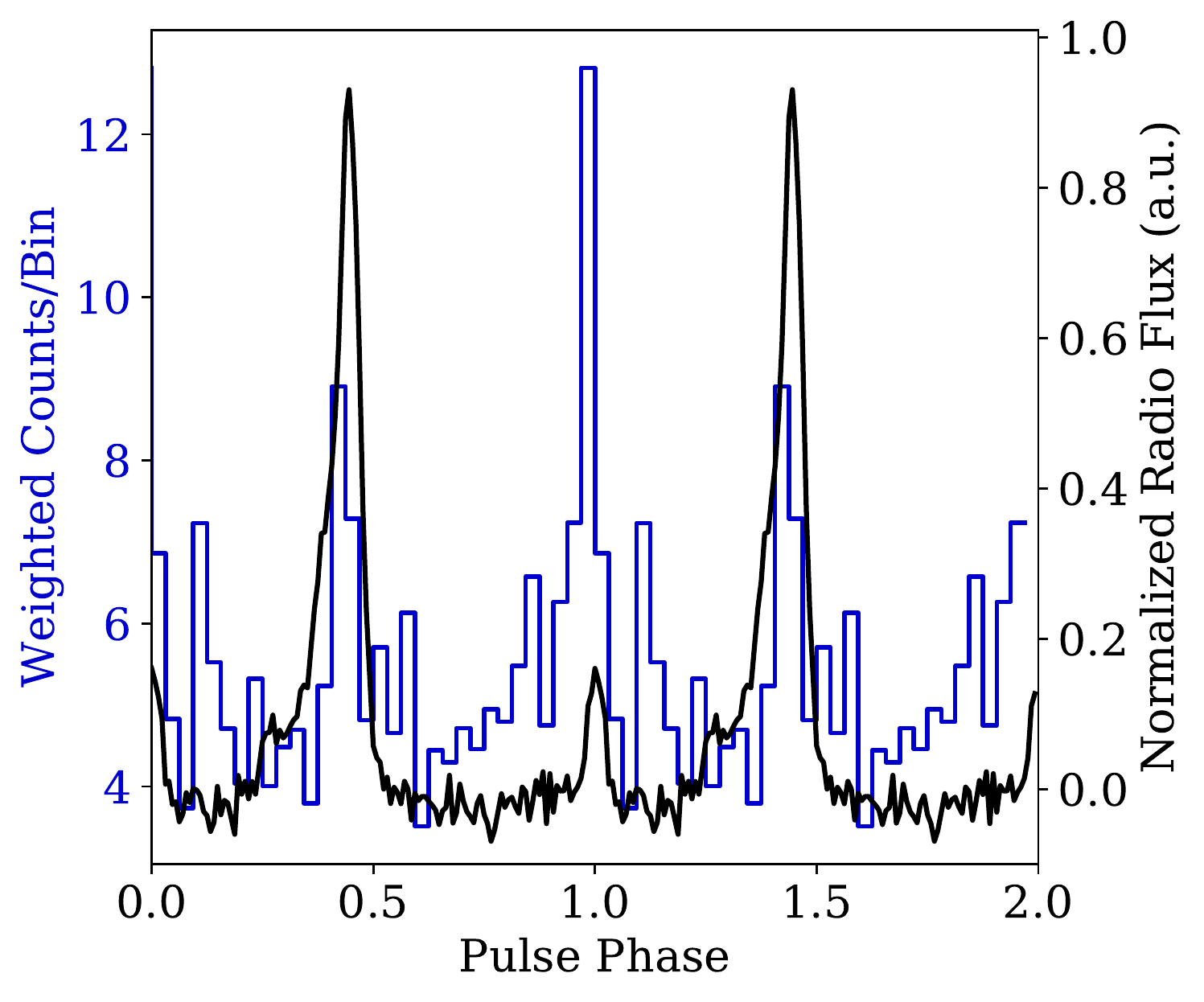}
    \caption{Radio (1.4\,GHz; black curve) and gamma-ray (0.1--300\,GeV; blue curve) pulse profiles for PSR J1921+1929. For clarity, two full rotational phases are shown for the profiles. Phase-alignment was obtained by folding photons using the same ephemeris, reference epoch and template from the radio analysis.}
    \label{fig:gamma_prof}
\end{figure}

All photons from MJDs 56000 to 58500 with reconstructed directions within 2$^{\circ}$ of the radio timing position of PSR J1921+1929 that have energies in the 0.1--300\,GeV range were retrieved, filtered and barycentered using the event selection criteria recommended by the \textit{Fermi} Science Support Center\footnote{\url{https://fermi.gsfc.nasa.gov/ssc/data/analysis/scitools/overview.html}}. We rejected photons with weights smaller than 0.02 and phase-folded using the radio timing ephemeris presented in Table \ref{tab:longtable1}. The resulting detection has a H-test value \citep{db10} of 129, corresponding to a 10$\sigma$ significance. The phase-aligned radio and $\gamma$-ray pulse profiles are shown in Figure \ref{fig:gamma_prof}.Both profiles have aligned primary and a secondary peaks, but the primary/secondary $\gamma$-ray peak aligns with the secondary/primary radio peak. 
%The first $\gamma$ peak leads the radio peak by a rotational phase of 0.625. PSR J1921+1929 exhibits two main peaks (separated by a phase of 0.47), and a third, weaker peak trails the primary. 
%Like the majority of $\gamma$-ray MSPs \citep{fermi13}, PSR J1921+1929 exhibits a double-peaked profile at high energies, further supporting the reality of the pulsations.  

The other pulsars reported in this work all have expected $\gamma$-ray flux $\sqrt{\dot{E}}/(4\pi D^2)$ \citep{fermi13} below 8\,$\pm$\,3\,$\times$\,10$^{14}$\,erg\,s$^{-1}$\,kpc$^{-2}$. Furthermore, they are located within 3$^\circ$ from the Galactic plane where the high background of diffuse emission further makes detection difficult. Photons within 2$^\circ$ of the positions of the other pulsars were extracted, filtered and phase-folded, but we did not detect any significant pulsations. Given the vast majority of known MSPs with estimated $\gamma$-ray flux $\sqrt{\dot{E}}/(4\pi D^2)\,<\,10^{15}$\,erg\,s$^{-1}$\,kpc$^{-2}$ are not detected by \textit{Fermi} \citep{sbc+18}, not detecting $\gamma$-ray pulsations for the remaining MSPs is expected.

In summary, we identify $\gamma$-ray pulsations for PSR J1921+1929, spatially coincident with the \textit{Fermi} unassociated point source 3FGL J1921.6+1934.

\section{Discussion} \label{Galactic distribution}
Establishing a more complete census of the Galactic MSP population is key to understanding stellar and binary evolution in interacting systems. Currently, the known population is highly biased towards nearby sources ($D\,\lesssim$\,2\,kpc). Detecting distant MSPs allows us to better model the binary density distribution for different classes in terms of galactocentric radius and height. This not only contributes to expanding our knowledge of binary evolution, but can also help in designing surveys to target certain types of pulsars.
%and challenging, especially for fast MSPs since they are more vulnerable to propagation effects. 
%Over the past years, more than 50 of the Galactic MSPs have been discovered via targeted radio searches of $Fermi$ unassociated sources (e.g., \citealt{kjr+11,kcj+12,ckr+15}).

To date, 38 MSPs have been discovered by PALFA, including 29 that have been confirmed to be in binaries. Figure \ref{fig:DMvsP} shows that the DMs of PALFA-discovered MSPs are among the highest known, as already discussed by \cite{csl+12}, \cite{skl+15} and \cite{sab+16}. The average DM/$P$ ratio is 13\,pc\,cm$^{-3}$\,ms$^{-1}$ for PALFA MSPs, three times larger than the non-PALFA recycled pulsars. Furthermore, they account for half of the known MSPs with DM\,$>$\,150\,pc\,cm$^{-3}$. Considering only the most recycled pulsars ($P\,<$\,10\,ms) at those DMs, 70$\%$ were discovered by PALFA and eleven of the 15 most dispersed were found by the survey. As a result of Arecibo's exceptional sensitivity and the high-resolution observations at L-band, our sensitivity to recycled pulsars is one of the highest, and matches the detectability limits predicted by the radiometer equation \citep{lbh+15}. Consequently, PALFA's contribution to unveiling the true MSP population in the Galactic field is among the most significant in radio pulsar surveys.

\subsection{Spatial distribution} \label{galpop}
%This near-isotropy would be expected of a population that has reached dynamic equilibrium.
\begin{figure*}[t]
    \includegraphics[width=1.03\textwidth]{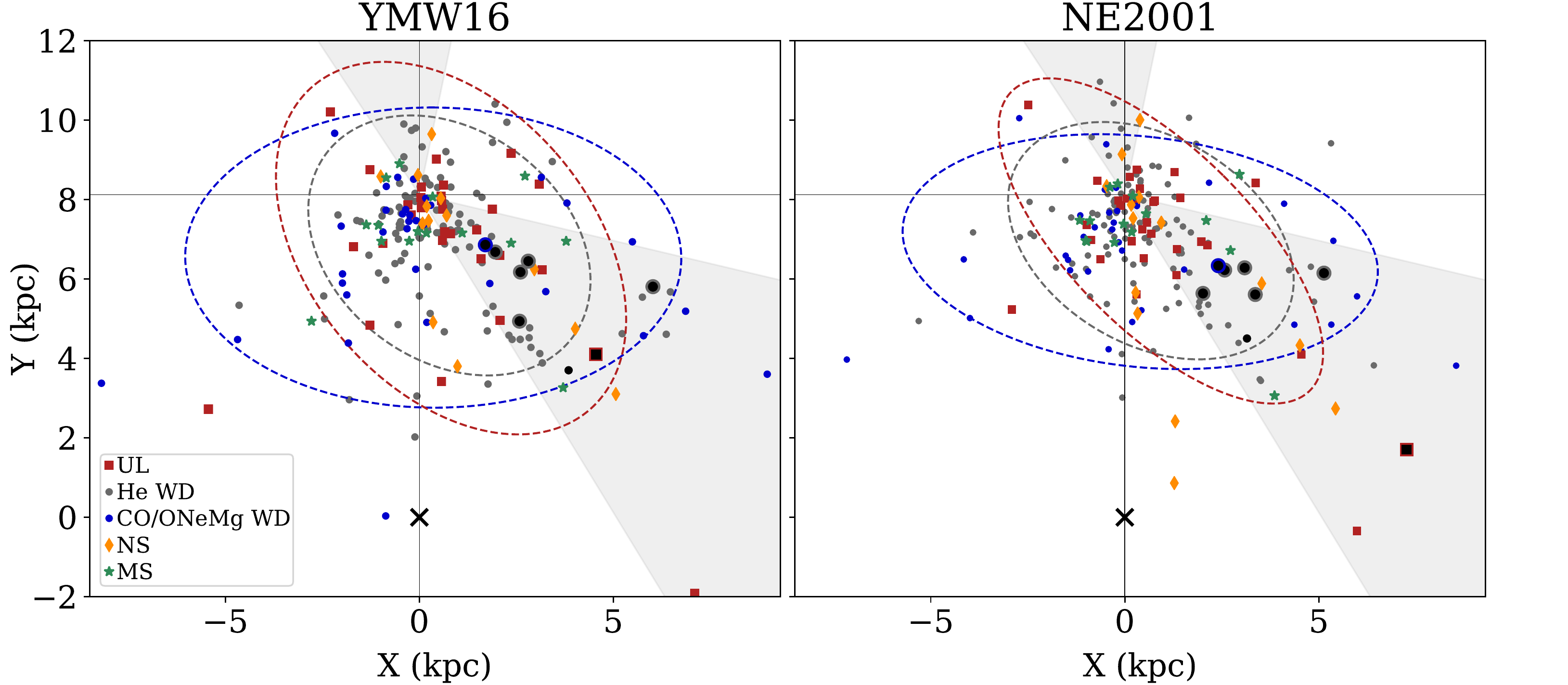}
    \caption{Spatial distribution of known binary MSPs in the X,Y,Z Cartesian coordinate system (based on the equatorial system) projected onto the X-Y Galactic plane using distance predictions from the YMW16 (\textit{left panel}) and NE2001 (\textit{right panel}) electron density models. Shaded sections are PALFA-covered regions, and the color-based classification for the binary type is the same as in Figure \ref{fig:binaries_params}. MSPs presented in this work are represented by black-filled circles. The black X marker corresponds to the Galactic Centre, and the Solar System Barycenter is located at the intersection of the two solid black lines. Dashed ellipses are the 2$\sigma$ fits to the X-Y Galactic coordinates for each binary class. Unlike the blue ellipse (CO/ONeMg WD companions), the semi-major axis of the red and grey ellipses (UL and He WD companions) are not parallel to the x-axis, indicative of an asymmetry about X\,=\,0\,kpc. This is due to the significant contribution of the PALFA survey to the known UL and He WD binaries. }
    \label{fig:xy_msps}
\end{figure*}
When conducting population syntheses, a radial scaling is typically assumed when assigning the birth distribution of the population. Increasing the spatial extent covered by observed MSPs is therefore necessary to determine the best radial distribution to use as initial conditions.

Using both YMW16 and NE2001 models, we map the Galactic distribution, projected onto the plane, of the observed population of radio MSPs in binaries. Results are shown in Figure \ref{fig:xy_msps}, where ellipses represent the 2$\sigma$ fits to the (X,Y) Galactic coordinates of the MSPs projected on the plane for the different binary classes. Looking at the location and distribution of points, one can see the observational bias toward nearby sources, especially for pulsars with ultra-light (UL) and He WD companions judging by the smaller ellipses. Those pulsars have undergone longer accretion phases and thus have shorter spin periods than pulsars with massive CO/ONeMg WDs. Short-$P$ pulsars are more difficult to find at larger distances in the Galactic plane since their signals are more strongly affected by propagation effects caused by the interstellar medium. We see however from Figure \ref{fig:xy_msps} that UL systems, which have the most rapidly spinning pulsars, are found at larger (projected) distances compared to those with He WDs. This is explained by the important increase in discoveries of high-$\dot{E}$ sources (i.e., very fast spinning pulsars such as those with UL companions) by \textit{Fermi}-LAT and targeted radio searches of \textit{Fermi} sources over the past years (e.g. \citealt{fermi09}, \citealt{fermi10}, \citealt{fermi13}, \citealt{fermi15}).

In addition to the bias towards nearby objects, a longitudinal asymmetry for binaries with low-mass companions is also observed. Unlike the expected latitudinal bias that results from the larger dispersive smearing and scattering near the plane, a binary-type dependence in the longitudinal distribution reflects different survey efficiencies. We interpret the skewness in (X,Y) position for He WD and UL binaries toward the PALFA-covered regions (shaded regions in Figure \ref{fig:xy_msps}) as a survey bias: PALFA outperforms most surveys in discovering distant, very recycled pulsars. The observed population is consequently not representative of the true field population, and this should be taken into consideration in population modeling. Having sensitive surveys conducted at high frequencies capable of finding fast MSPs in the Galactic plane is therefore important for accurately modeling the population. % SKA

\subsection{Galactic height} \label{galheight}
Kick velocities imparted to neutron stars following asymmetric supernova (SN) explosions strongly impact the Galactic height distribution of pulsars. In this event, overall momentum is conserved. But where for isolated pulsars the accelerated mass is only that of the neutron star, binary pulsars need to also drag along their companion. Due to this higher system mass, one theoretically expects the recoil velocity to be smaller for binary pulsars compared to isolated pulsars (e.g. \citealt{kh09}), and to decrease with increasing companion mass. The Galactic height of observed pulsars belonging to various populations are used as probes in testing predictions emerging from simulations. 
%Assessing the absolute Galactic height distributions of the various classes of MSP-binaries is thus necessary to evaluate models. 

\begin{figure*}[ht!]
  \centering
    \includegraphics[width=1.0\textwidth]{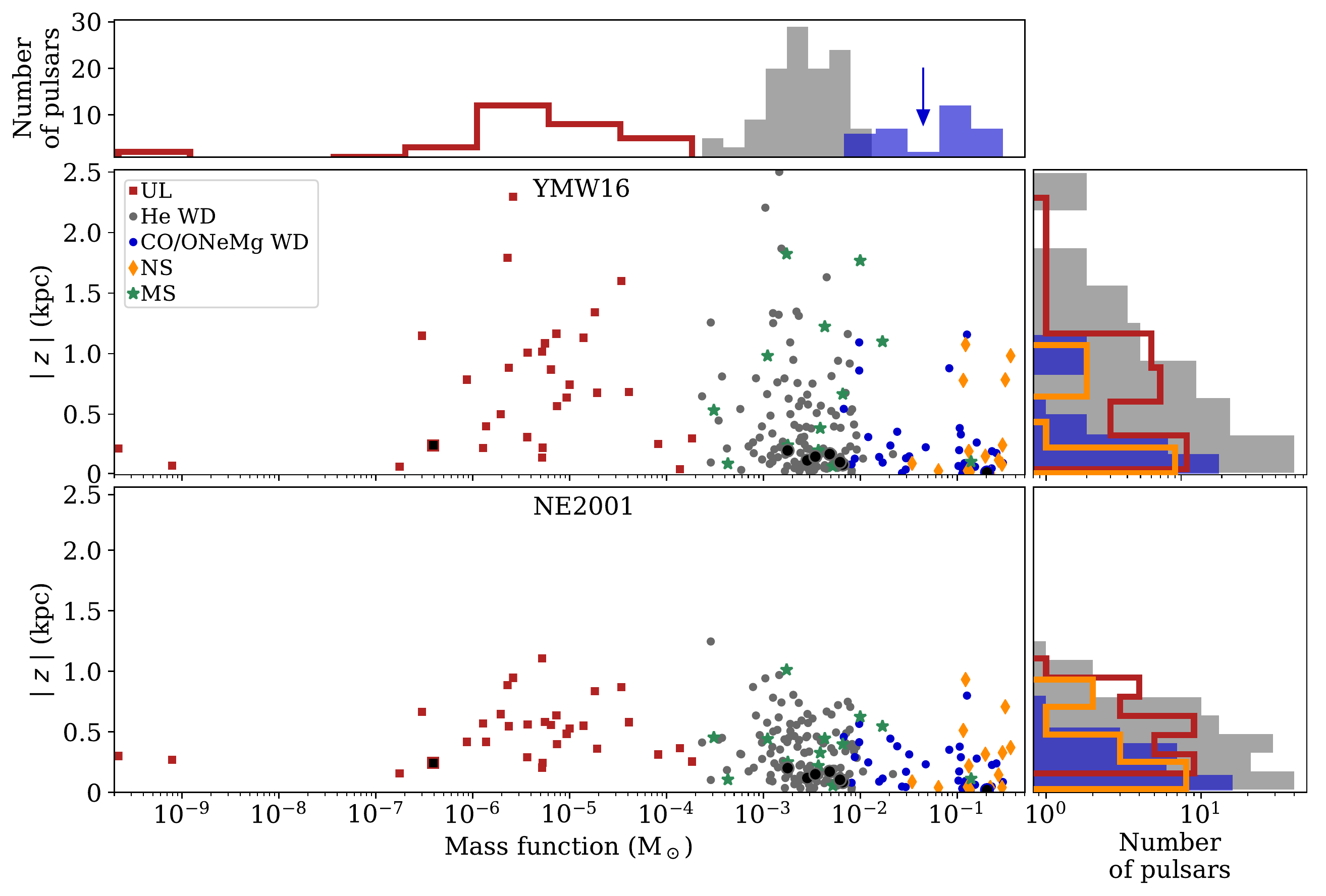}
    \caption{\textit{Bottom left panel}: Absolute Galactic height $| z |$ versus mass function of Galactic MSPs in binary systems. Open symbols correspond to PALFA-discoveries and black-filled symbols are the MSPs presented in this work. \textit{Top panel}: Mass function distributions for systems with UL, He WD and CO/ONeMg WD companions. The blue arrow points to the gap in mass functions of MSP-CO/ONeMg WD binaries we discuss in Section \ref{nswd}. To help visualize more clearly this gap, histograms for MSP-NS and MSP-MS mass functions are not plotted on this panel. \textit{Right panel}: Height distributions for those same binary categories. Overdensities of UL and He WD binaries at $| z |\,\sim$\,1\,kpc are the result of the increasing number of \textit{Fermi}-directed radio discoveries. Also for clarity, we do not plot the heights of MSP-MS systems.}
    \label{fig:zz_mf}
\end{figure*}

Additionally, performing such a comparison is relevant for estimating the contribution of pulsars to the unresolved diffuse $\gamma$-ray background, a long-standing open problem in astrophysics. Indeed, the $\gamma$-ray luminosity of a pulsar scales with spin-down power and is consequently greater for highly recycled pulsars with light companions (e.g. \citealt{fermi13}). Hence, assessing a representative height distribution for the different binary classes is highly motivated.
 
\cite{nbb+14} (hereafter N14) investigated the possible correlation between the mass function of MSP binaries in the Galactic field and their vertical distance from the Galactic plane. Their primary conclusions were that (1) more massive systems are found closer to the Galactic plane and (2) there is a larger scatter in the height distribution of lighter systems. 

We reproduce this analysis using an updated and larger sample of 214 MSP binaries\footnote{Again, using the following definition of recycled pulsars: $P\,<$\,100\,ms and $\dot{P}\,<\,1\,\times\,10^{-17}$.} in the field (versus 164 in N14), taken from the ATNF Catalogue and the Galactic MSP Catalog\footnote{\url{http://astro.phys.wvu.edu/GalacticMSPs/GalacticMSPs.txt}}. Similarly to N14, we adopt the \cite{tlk12} prescription to classify systems with unspecified companion type. 

Distance estimates were calculated with both the NE2001 and YMW16 electron density models, and we compare the resulting projected heights. For every binary class, YMW16 produces larger absolute mean heights, $| z_{mean} |$, and standard deviations in absolute heights $\sigma$ (see Table \ref{tab:height2}) than NE2001. We note however that YMW16 predicts distances larger than NE2001 for only 55\% of the pulsars, so the larger $| z_{mean} |$ values are not a consequence of systematic distance overestimations compared to NE2001. While NE2001 predicts that all pulsars are within 1.25\,kpc from the Galactic plane, some systems could be found at heights as large as 2.50\,kpc according to YMW16 predictions. In the latter model, pulsars having $| z | > $ 1.0\,kpc the distant, high-latitude ($|b|>\,15^{\circ}$) systems. In a recent study that used new parallax measurements, \cite{dgb+19} demonstrated that while YMW16 performs moderately better than NE2001 at high latitudes, predictions that were largely inconsistent with parallax measurements were mostly high-latitude objects. We therefore emphasize that interpretations from this analysis are subjected to significant uncertainties and should be considered with caution. Results for each binary classes are discussed below.
\input{table_heights2.tex}

\subsubsection{MSP-WD binaries} \label{nswd}
When considering systems with WD companions, our results agree with and strengthen conclusion (2) of N14: lighter MSP-WD systems show more scatter in their Galactic height distribution than the massive ones. This is true for both electron density models. Compared to pulsars with CO/ONeMg WDs, those with UL companions have estimated $| z_{mean} |$ larger by factors of two and three based on the NE2001 and YMW16 models, respectively. 
%We find that UL and He WD binaries have similar standard deviations in $| z |$ ($\sigma$\,=\,0.52\,kpc), two times that of MSPs with CO or ONeMg WD, while N14 reported $\sigma$\,=\,0.22\,kpc for UL and 0.26\,kpc for He WD companions versus 0.15\,kpc for massive WDs. 

As mentioned in N14, while our ability to detect rapidly rotating MSPs (i.e., the more recycled, lighter systems) is strongly reduced as we search regions of high electron density close to the disk, the detectability of less-recycled pulsars with massive WDs is not as height-dependent since the typical timescales of pulse dispersion/scattering are smaller relative to the pulse widths of these longer-spin-period pulsars. MSPs with CO/ONeMg WDs are therefore easier to detect. Consequently, the observed lack of heavy systems far from the Galactic plane seems genuine and the mean absolute distance from the plane $| z_{mean} |$ (0.25\,kpc for YMW16 and 0.21\,kpc for NE2001) is most probably representative of the true population for this class. 

As a result of stronger observational biases, the measured $| z_{mean} |$ (0.72\,kpc for YMW16 and 0.51\,kpc for NE2001) of the lighter MSP-UL systems is likely overestimated. Evidence for this is the $| z |$ distribution in Figure \ref{fig:zz_mf}, which reveals an overdensity of objects at an absolute height of $\sim$\,0.8\,kpc. The majority of those MSPs have been recently discovered via \textit{Fermi}-directed searches (\citealp{rap+12}). Considering the higher detection threshold at low Galactic latitude of \textit{Fermi}-LAT due to the higher background diffuse emission \citep{fermi15}, we can extrapolate from those overdensities and predict a larger population near the Galactic plane.  

N14 noted that pulsar ages could influence the observed scatter in heights: fully recycled pulsars with lower-mass companions are generally older than those with massive companions, resulting in lighter systems having more time to move away from the Galactic plane. We know however that in- and out-of-the-plane oscillatory motions of neutron stars (having velocities smaller than the Galaxy's escape velocity) in response to the Galactic gravitational field have periods of the order of $10^7$\,yrs \citep{cc98}, much shorter than the age of MSPs. Current scale heights could however be affected by the age of the MSP due to the different scale heights at birth associated to system progenitors as well as time-scales associated to gravitational diffusion processes \citep{cc98}. Consequently, the larger scatter in scale heights for lower-mass-function systems is most likely a result of correlations between binary mass function, systemic recoil velocity, supernova processes, birth scale height distribution and Galactic diffusion.

We also notice a mild deficiency of systems with CO or ONeMg WDs having mass function between $\sim$\,4\,$\times\,10^{-2}$ and 9\,$\times\,10^{-2}\,M_\odot$ in the top panel of Figure \ref{fig:zz_mf} (indicated by a blue arrow), resulting in two possible sub-populations. Approximately 40$\%$ of our sample of MSP-CO/ONeMg WD binaries falls below this apparent gap. To determine if the observed mass functions could originate from a uniform distribution, we generated trial samples and performed a two-sided Kolmogorov-Smirnov (KS) tests. The samples shared the same mass function interval as the observed sample and were of the same size. To avoid fluctuations due to small-number statistics, we generated 10 000 trial samples. The average probability that the observed and trial samples share a common parent distribution is 3\%, suggesting that the MSP-CO/ONeMg mass functions are not drawn from a uniform distribution.
%A Kolmogorov-Smirnov test indicates that the probability that the two samples are drawn from the same continuous distribution is $\sim\,10^{-8}$, suggesting that they represent two different populations.} 

The known orbital period gap at $P_b\,\sim\,25-50$\,days (e.g., \citealt{t96,tkr00,hwh+18}) that possibly arises from a bifurcation mechanism leading to divergent evolutionary paths \citep{t96} does not explain this apparent population splitting: members of these sub-populations are found on both sides of this $P_b$ gap. We note here that two of the pulsars presented in this work, PSRs J1921+1929 and J1932+1756, fall within this period gap. Furthermore, the fraction of higher-mass-function systems seems too high to be explained by systems where the companions are the more massive ONeMg WDs, which are rarer than the lighter CO WDs. An alternative interpretation is an asymmetry in mass ratios, possibly reflecting the asymmetric distribution of the NS masses found by previous studies \citep{vrh+11,opn+12,kky+13,ato+16}. Further discoveries will help in confirming the existence of this possible gap within MSP-massive WD binaries.

\subsubsection{MSP-NS binaries} \label{nsns}
Because DNSs endure two SN kicks rather than one, DNS heights are not interpreted in the same manner as those involving WD companions. From Figure \ref{fig:zz_mf}, we note that there seems to be a bimodality in the Galactic heights for the YMW16 estimations, uncorrelated with the mass function, in which ten DNSs have $| z |\,<\,0.25$\,kpc and the remaining four have $| z |$ between 0.75 and 1.1\,kpc. This potential separation in the observed population is however not as significant in the NE2001 predictions. If real, this could be related to the bimodal natal kick distributions that arise from the different SN mechanisms giving birth to the NSs (e.g. \citealt{spr10,afk+15,bp16,vns+18}).
  
Prior to the second SN, all future DNS systems are completely circularized, as a result of the accretion process. When the second SN occurs, the magnitude and direction of the associated kick and the SN mass loss determine the post-SN orbital eccentricity and the system's velocity relative to the Galaxy: the larger the kick, the larger (on average) will be the post-SN orbital eccentricities and motions relative to the Galaxy. A positive correlation between these quantities has indeed been detected \citep{tkf+17}.

Thus the systems with low orbital eccentricity ($e\,<\,0.15$) and low peculiar velocity, which likely had a small kick (eg. PSRs J0453+1559; \citealt{msf+17}, J0737$-$3039A/B; \citealt{bap+03,lbk+04}, 
J1906+0746; \citealt{vks+15}, J1913+1102; \citealt{lfa+16}, J1946+2052; \citealt{sfc+18}) should always be located near the Galactic plane; this is indeed observed: their maximum Galactic height is about 0.3\,kpc. Broader distributions of orbital periods, eccentricities and systemic velocities are expected for systems with large-kick second SNe \citep{dpp05}. For $e\,>\,0.15$, the range of Galactic heights implied by the YMW16 model increases to 1.2\,kpc; all high-altitude ($z\,>\,0.7$\,kpc) systems (PSRs B1534+12; \citealt{w91}, J1518+4904; \citealt{nst96}, J1411+2551; \citealt{msf+17}, J1757$-$1854; \citealt{cck+18}) have $e >$\,0.15. Those pulsars also have the highest $| z |$ values implied by NE2001, but their range in $| z |$ only extends from 0.37 to 0.93\,kpc away from the plane. We note that a distance measurement based on the orbital decay of PSR B1534+12 \citep{sac+98,fst+14} was used as a predictor while constructing both electron density models, therefore the measured Galactic height for this source is DM-independent.

It will be important to see if future discoveries confirm the significance of this potential gap in DNS heights. This type of study will enable increasingly accurate models of massive stellar binary evolution and SN kick distributions.
%kramer 2006: proper motion low velocity - double psr 0737 
%B1534+12 -> Kalogera, last few years, jeff 2015,  evolutionary channel for evolution of DNS
%double ns galactic - vigna-gomez 2018 
\subsubsection{MSP-MS binaries} \label{nsms}
Galactic, recycled pulsars in orbit with low-mass, hydrogen-rich (likely non-degenerate) stars are difficult to detect. This is a consequence of radio eclipses caused by the stellar material surrounding the system. Those are referred to as ``redback'' systems (\citealt{r13}), and only 14 have been confirmed as such (see recent work by \citealt{ssc+18}). The 15 MSP-MS binaries in our sample have MSPs with $P <$\,8\,ms and DM\,$<$\,70\,pc\,cm$^{-3}$ except PSR J1903+0327, a unique PALFA discovery. The latter system has a DM of 298\,pc\,cm$^{-3}$ and is in an eccentric orbit ($e$\,=\,0.44, \citealt{crl+08}) with a star of spectral type between F5V and G0V \citep{fbw+11,ksf+12}. It is the only member of our sample that is not a redback. 

From Figure \ref{fig:zz_mf} and Table \ref{tab:height2}, we see that systems with MS stars have mass functions, $| z_{mean} |$ and $\sigma$ values similar to those with WDs ($M_{WD}\,\gtrsim\,0.1\,M_\odot$): a large scatter in both their heights and mass functions. Similarly to MSPs with UL/He WD companions, their short spin periods make them harder to detect because of interstellar dispersion and scattering. In recent years, redbacks have been largely discovered via targeted searches of \textit{Fermi} unassociated sources (e.g., \citealt{hrm+11,cck+16}), which also has a reduced detection capability close to the disk \citep{fermi15}. Thus, the same observational bias against near-the-plane discoveries applies and we conclude that $| z_{mean} |$ is likely overestimated for this binary class as well. 

\section{Summary and Conclusion} \label{summary}
Timing solutions for eight new PALFA-discovered MSPs in binary systems were presented in this work. All follow-up observations were conducted at 1.4\,GHz at AO and JBO. 

Among the discoveries are (1) a black widow pulsar (PSR J1928+1245) for which we do not detect eclipses, (2) a new $\gamma$-ray MSP (PSR J1921+1929) that we associate with a \textit{Fermi} point source, and (3) an intermediate-mass binary pulsar (PSR J1932+1756) in a low-eccentricity orbit with a massive WD. Seven of the new discoveries are very fast-spinning ($P\,<$\,6\,ms), binary MSPs deep in the Galactic plane, which are particularly difficult to detect as a result of pulse dispersion and scattering by the dense interstellar medium. Most systems are in nearly circular orbits in which the most probable companions are He WDs. Except for $\gamma$-ray pulsations from PSR J1921+1929, we do not detect any other multiwavelength counterparts or possible associations to the pulsars. 

In this work, we also analyzed the Galactic distribution of 214 recycled pulsars belonging to different binary classes. We see that MSPs with massive WDs show less scatter in their absolute Galactic heights compared to lighter systems, in agreement with results from \cite{nbb+14}. MSP-CO/ONeMg WD binaries being less vulnerable to observational biases, we believe that our measured average heights, $| z_{mean} |$, of 0.25 and 0.21\,kpc for the YMW16 and NE2001 electron density models, respectively, are more representative of the true population than MSPs in orbit with lighter WDs.

We identify a longitudinal bias for finding the most recycled MSP-binaries towards PALFA-covered regions, and an overdensity of such objects at $| z |\,\sim\,$0.8\,kpc, which reflects the increased number of discoveries from \textit{Fermi}-directed radio searches. Those are indications that we are starting to unveil a more representative population near the plane. However, PALFA only surveys regions at $\lvert b \lvert\,< 5^{\circ}$. This itself represents an important selection bias. Without searching regions of higher Galactic latitude with surveys having PALFA-like sensitivity, a sample more representative of the true population cannot be produced. 

A potential gap in the mass function of systems with massive WDs at $\sim$\,4\,$\times\,10^{-2}$ to 9\,$\times\,10^{-2}$\,$M_\odot$ is found, potentially reflecting a bimodal distribution in  binary mass ratios. 

Additionally, distances estimated with the YMW16 model suggest a potential gap, uncorrelated with eccentricity, in the absolute heights of the 14 known DNSs in the Galactic field at approximately 250\,$\lesssim\, | z | \,\lesssim$\,750\,pc. If this gap persists as more systems are discovered, we could gain new information on the supernova mechanisms producing those DNSs. 

The new PALFA discoveries presented in this work belong to this population of faint, high-DM MSPs, and the non-detection of high-energy pulsations for all but one of those new pulsars indicates that PALFA is complementary to the ongoing survey with \textit{Fermi}. Further discoveries from the survey will continue to unveil a more complete census of the Galactic MSP population in the plane.  %, enabling unique and exciting studies.  

\section*{Acknowledgements}
The Arecibo Observatory is operated by SRI International under a cooperative agreement with the National Science Foundation (AST-1100968), and in alliance with Ana G.M\'{e}ndez-Universidad Metropolitana, and the Universities Space Research Association. The National Radio Astronomy Observatory is a facility of the National Science Foundation operated under cooperative agreement by Associated Universities, Inc. The CyberSKA project was funded by a CANARIE NEP-2 grant. The processing of survey data was made on the supercomputer Guillimin at McGill University, managed by Calcul Qu\'{e}bec and Compute Canada. The operation of this supercomputer is funded by the Canada  Foundation  for  Innovation (CFI),  NanoQu\'{e}bec, RMGA and the Fonds de recherche du Qu\'{e}bec$-$Nature et technologies (FRQ-NT). 

EP is a Vanier Canada Graduate Scholar. VMK acknowledges support from an NSERC Discovery grant and Herzberg Award, the Canada Research Chairs program, the Canadian Institute for Advanced Research, and FRQ-NT. AB, FC, IHS, JMC, SC and SMR are members of the NANOGrav Physics Frontiers Center, which is supported by the National Science Foundation award number 1430284. JSD was supported by the NASA Fermi program. JvL acknowledges funding from European Research Council grant n. 617199 (`ALERT'), and from Netherlands Organisation for Scientific Research (NWO) Vici grant 639.043.815 (`ARGO'). MAM is supported by NSF award OIA-1458952. SMR is a CIFAR Fellow. P.S. is supported by a DRAO Covington Fellowship from the National Research Council Canada. Pulsar research at UBC is supported by an NSERC Discovery Grant and by the Canadian Institute for Advanced Research. Pulsar research at the University of Manchester is supported by a Consolidated Grant from STFC. WWZ is supported by the National Key RD Program of China No. 2017YFA0402600, the Strategic Priority Research Program of the Chinese Academy of Sciences Grant No. XDB23000000, and by the National Natural Science Foundation of China under grant No. 11690024, 11743002, 11873067.  \\ \\

\bibliography{PALFA-MSP-Timing-v2}
\bibliographystyle{aasjournal}
\end{document}

%% file: table-timing-1.tex
\multicolumn{1}{l}{\textbf{ }} & \multicolumn{1}{c}{J1906+0454} & \multicolumn{1}{c}{J1913+0618} & \multicolumn{1}{c}{J1921+1929} & \multicolumn{1}{c}{J1928+1245} \\ \hline 
\hline \hline
Right ascension (J2000)                                     & 19:06:47.13970(7) & 19:13:17.8786(2)  & 19:21:23.35070(4) & 19:28:45.39360(6)     \\
Declination (J2000)                                         & +04:54:09.042(2)   & +06:18:08.277(5)   & +19:29:22.299(1)   & +12:45:53.374(3)       \\
Proper motion in right ascension (mas\,yr$^{-1}$)		    & $-$6(1)           & $-$9(2)           & $-$3.2(8)         & $<$3                  \\
Proper motion in declination (mas\,yr$^{-1}$)   		    & $-$14(3)          & $<$12             & $-$11(1)          & $<$15                 \\
Total proper motion, $\mu_T$ (mas\,yr$^{-1}$)               & 15(3)             & 9(2)              & 12(1)             & $<$15                  \\
Parallax, $\pi$ (mas)                                       & $<$12             & $<$17             & $<$9              & $<$16                \\
Spin frequency, $\nu$ (s$^{-1}$)                            & 480.01267196822(5)& 198.93129209203(3)& 377.88018632311(2)& 330.94979452682(6)	\\
Spin frequency derivative, $\dot{\nu}_{obs}$ ($ 10^{-15}$\,s$^{-2}$)& $-$0.687(3)& $-$0.3804(8)      & $-$5.4542(7)     & $-$1.84(1)            \\
Dispersion Measure, DM (pc\,cm$^{-3}$)             		    & 142.793(2)        & 155.987(3)        & 64.7632(9)        & 179.233(1)            \\
Orbital period, $P_b$ (days)                                & 7.04187028(1)     & 67.7439374(2)     & 39.64926375(2)    & 0.1366347269(8)       \\
Projected semi-major axis, $x=a$sin$i/c$ (lt-s)             & 4.350215(2)       & 32.742919(2)      & 20.8474224(7)     & 0.018951(1)           \\
Time of periastron, $T_{0}$ (MJD)                           & -                 & 57721.746(3)      & -                 & -                     \\
Time of ascending node, $T_{asc}$ (MJD)                     & 58101.4031477(5)  & -                 & 57602.9409583(2)  & 57949.020215(3)       \\
Eccentricity, $e$ ($ 10^{-4}$)                              & -                 & 7.053(1)          & -                 & -                     \\
1$^{\rm st}$ Laplace parameter, $\epsilon_1$\,=\,$e$ sin $\omega$ ($ 10^{-4}$) & $<$0.03   & -      & 0.8093(5)         & $<$5                  \\
2$^{\rm nd}$ Laplace parameter, $\epsilon_2$\,=\,$e$ cos $\omega$ ($ 10^{-4}$) & $<$0.02   & -      & $-$1.9182(6)      & $<$4                  \\
Longitude of periastron, $\omega$ (deg)                     & -                 & 38.65(2)          & -                 & -                     \\ \\

Galactic longitude, $l$ (deg)			  	       	        & 39.045            & 41.034            & 53.619            & 48.534                 \\
Galactic latitude, $b$ (deg) 			  	   	            & $-$1.167          & $-$1.960          & 2.453             & $-$2.290               \\
Period, $P$ (ms) 						  	                & 2.0832783349232(2)& 5.0268612317532(8)& 2.6463414494692(1)& 3.0216063479651(5)     \\
Period derivative, $\dot{P}_{obs}$\tnote{a}\hspace{0.2cm}($ 10^{-20}$) 	& 0.298(1)          & 0.961(2)          & 3.8196(5)         & 0.168(1)               \\
Mass function (10$^{-3}\,M_\odot$)                          & 1.782539(2)       & 8.212853(2)       & 6.1882748(6)      & 0.0003914(1)           \\
Minimum companion mass ($M_\odot$)        	      		    & 0.163             & 0.286             & 0.257             & 0.009                  \\
Median companion mass ($M_\odot$)         	        	    & 0.191             & 0.337             & 0.302             & 0.011                  \\
Surface dipolar magnetic field, $B$\tnote{b}\hspace{0.2cm}($ 10^8$\,G) 	& 0.8               & 1.8               & 3.2               & 0.7                    \\
Spin down luminosity, $\dot{E}$\tnote{b}\hspace{0.2cm}($ 10^{33}$\,erg\,s$^{-1}$) & 13      & 2.0               & 81                & 2.4                    \\
Characteristic age, $\tau_c$\tnote{b}\hspace{0.2cm}(Gyr) 	& 11                & 12                & 1.1               & 28                     \\
DM distance, $D$\tnote{c}\hspace{0.2cm}(kpc)                & 4.1               & 5.9               & 2.4               & 6.1                    \\
Pulse width, W50\tnote{d}\hspace{0.2cm}(ms)                 & 0.20              & 0.47              & 0.17              & 0.28                   \\
Rotation measure, RM (rad m$^{-2}$)                         & -                 & -                 & 121(22)           & 60(17)                 \\
Mean flux density, $S_{1400}$\tnote{e}\hspace{0.2cm}(mJy)   & 0.043(4)          & 0.057(7)          & 0.12(2)           & 0.080(6)               \\
Mean luminosity, $L_{1400}$\tnote{f}\hspace{0.2cm}(mJy\,kpc$^2$)& 9                 & 25                & 72                & 26                     \\ \\    

Follow-up site                                              & AO               & AO                 & AO                & AO            \\
Discovery observation date (MJD)                            & 57258            & 56817              & 57573             & 57268         \\
Reference epoch (MJD)                                       & 58100.0          & 57696.0            & 58137.0           & 57974.0       \\
First TOA (MJD)                                             & 57520.34         & 56935.93           & 57595.19          & 57268.02      \\
Last TOA (MJD) 											    & 58680.17         & 58456.78           & 58680.22          & 58680.21      \\
Number of TOAs 												& 143              & 131                & 123               & 135           \\           
Binary model                                                & ELL1             & DD                 & ELL1              & ELL1          \\
EFAC\tnote{g}                                                        & 1.1              & 1.2                & 1                 & 1             \\
rms post-fit residuals ($\mu$s) 				            & 12.09            & 19.78              & 4.43              & 10.35         \\ 
Daily-averaged rms post-fit residuals ($\mu$s)              & 8.42             & 6.06               & 2.40              & 2.63          \\ 
Reduced $\chi^2$                                            & 1.03             & 1.01               & 1.03              & 1.07          \\\hline \hline
\insertTableNotes 

%% file: table-timing-2.tex
\multicolumn{1}{l}{\textbf{ }} & \multicolumn{1}{c}{J1930+2441} & \multicolumn{1}{c}{J1932+1756} & \multicolumn{1}{c}{J1937+1658} & \multicolumn{1}{c}{J2010+3051} \\ \hline 
\hline \hline
Right ascension (J2000)                                     & 19:30:45.39844(6)     & 19:32:07.2277(5)     & 19:37:24.90851(8)  & 20:10:03.9542(1)     \\           
Declination (J2000)                                         & +24:41:01.564(1)       & +17:56:17.31(1)       & +16:58:42.553(2)    & +30:51:18.892(1)       \\               
Proper motion in right ascension (mas\,yr$^{-1}$)		    & $<$4                  & $-$13(5)             & $-$1.8(7)          & $-$10.3(9)              \\ 
Proper motion in declination (mas\,yr$^{-1}$)   		    & $-$10(3)              & $<$32                & $-$5(1)            & $-$3(1)               \\
Total proper motion, $\mu_T$ (mas\,yr$^{-1}$)               & 11(3)                 & 15(7)                & 5(1)               & 10.8(9)                 \\  
Parallax, $\pi$ (mas)                                       & 16(6)                 & $<$66                & $<$ 10             & $<$ 19                \\
Spin frequency, $\nu$ (s$^{-1}$)                            & 173.38814681699(2)    & 23.90555115685(1)    & 252.67407210877(1) & 207.62936108328(2)    \\                	
Spin frequency derivative, $\dot{\nu}_{obs}$ ($ 10^{-15}$\,s$^{-2}$)& $-$0.260(2)   & $-$0.2470(3)         & $-$0.8429(3)       & $-$0.2098(7)          \\              
Dispersion Measure, DM (pc\,cm$^{-3}$)             		    & 69.572(1)             & 53.183(9)            & 105.840(1)         & 133.756(2)            \\
Orbital period, $P_b$ (days)                                & 76.4059750(2)         & 41.5079763(1)        & 17.30610529(2)     & 23.35889575(2)        \\
Projected semi-major axis, $x=a$sin$i/c$ (lt-s)             & 29.748786(2)          & 68.85890(3)          & 9.2582711(9)       & 12.051767(1)          \\
Time of periastron, $T_{0}$ (MJD)                           & -                     & 57506.691(7)         & -                  & -                     \\
Time of ascending node, $T_{asc}$ (MJD)                     & 58162.4048255(7)      & -                    & 57750.5735047(9)   & 57844.9713383(5)      \\
Eccentricity, $e$ ($ 10^{-4}$)                              & -                     & 2.468(4)             & -                  & -                     \\
1$^{\rm st}$ Laplace parameter, $\epsilon_1$\,=\,$e$ sin $\omega$ ($ 10^{-4}$)& 1.210(1) & -               & $-$0.374(2)        & $-$0.056(2)           \\
2$^{\rm nd}$ Laplace parameter, $\epsilon_2$\,=\,$e$ cos $\omega$ ($ 10^{-4}$)& 1.2920(9)& -               & 0.838(2)           & $-$0.142(2)           \\
Longitude of periastron, $\omega$ (deg)                     & -                    & $-$55.50(6)           & -                  & -                     \\ \\
Galactic longitude, $l$ (deg)			  	       	        & 59.228                & 53.464               & 53.237             & 68.967                \\    
Galactic latitude, $b$ (deg) 			  	   	            & 3.002                 & $-$0.520             & $-$2.090           &  $-$1.326             \\             
Period, $P$ (ms) 						  	                & 5.7674069326981(6)    & 41.83128819908(2)    & 3.9576676453354(2) & 4.8162745133088(4)    \\             
Period derivative, $\dot{P}_{obs}$\tnote{a}\hspace{0.2cm}($ 10^{-20}$) 	& 0.866(6)              & 43.22(5)             & 1.3203(4)          & 0.487(2)              \\
Mass function (10$^{-3}\,M_\odot$)                          & 4.8421235(8)          & 203.4697(2)          & 2.8449441(8)       & 3.444533(1)           \\
Minimum companion mass ($M_\odot$)        	      		    & 0.235                 & 1.077                & 0.193              & 0.207                 \\
Median companion mass ($M_\odot$)         	        	    & 0.276                 & 1.325                & 0.226              & 0.243                 \\
Surface dipolar magnetic field, $B$\tnote{b}\hspace{0.2cm}($ 10^8$\,G) 	& 1.8                   & 40                   & 2.3                & 0.8                   \\
Spin down luminosity, $\dot{E}$\tnote{b}\hspace{0.2cm}($ 10^{33}$\,erg\,s$^{-1}$)   & 1.2        & 2.0                  & 8.5                & 0.4                   \\
Characteristic age, $\tau_c$\tnote{b}\hspace{0.2cm}(Gyr) 	   	   	   	& 16                    & 1.8                  & 4.7                & 66                    \\
DM distance, $D$\tnote{c} \hspace{0.2cm}(kpc)                           	& 3.3                   & 2.1                  & 3.3                & 6.5                   \\
Pulse width, W50\tnote{d}\hspace{0.2cm}(ms)                              & 0.32                  & 1.31                 & 0.40               & 0.23                  \\
Rotation measure, RM (rad m$^{-2}$)                         & -                     & -                    & -                  & -64(15)                \\
Mean flux density, $S_{1400}$\tnote{e} \hspace{0.2cm}(mJy)                &  0.084(2)             &  0.045(4)            & 0.09(5)            & 0.139(6)              \\
Mean luminosity, $L_{1400}$\tnote{f} \hspace{0.2cm}(mJy\,kpc$^2$)         &  11                   & 2                    & 12                 & 74                    \\ \\
Follow-up site                                              & AO+JBO                & AO+JBO               & AO                 & AO+JBO                \\
Discovery observation date (MJD)                            & 57701                 & 56537                & 56816              & 57209                \\    
Reference epoch (MJD)                                       & 58177.0               & 57516.0              & 57516.0            & 57949.0               \\    
First TOA (MJD)                                             & 57701.88              & 56537.03             & 56816.27           & 57218.800             \\
Last TOA (MJD) 											    & 58653.28              & 58496.67             & 58653.25           & 58680.24              \\            
Number of TOAs 												& 162                   & 216                  & 153                & 225                   \\
Binary model                                                & ELL1                  & DD                   & ELL1               & ELL1                  \\        
EFAC\tnote{g}                                                        & 1                     & 1                    & 1                  & 1.25                  \\ 
rms post-fit residuals ($\mu$s) 				            & 8.88                  & 51.92                & 6.88               & 13.31                 \\
Daily-averaged rms post-fit residuals ($\mu$s)              & 12.03                 & 22.09                & 1.74               & 7.09                  \\ 
Reduced $\chi^2$                                            & 0.74                  & 1.04                 & 1.07               & 1.00                  \\\hline \hline
\insertTableNotes 

%% file: table_mu_d_v.tex
%\begin{document}
\begin{table}[h]
  \centering
  \caption{Best-fit values for the total proper motion $\mu_{T,MC}$, distance $D$, transverse velocity $v_T$ and corresponding 1-$\sigma$ uncertainty produced by the Monte Carlo method described in Section \ref{fig:ppdot}. Initial uncertainties on the distance were set to 25$\%$. } 
  \setlength{\tabcolsep}{4.0mm}
  \begin{tabular}{c c c c}
\hline
PSR & $\mu_{T,MC}$ & $D$ &$v_T$  \\
    & (mas\,yr$^{-1}$)&(kpc)  & (km\,s$^{-1}$) \\                 
\hline \hline
J1906+0454  & 12.4$_{-1.1}^{+1.4}$      & 3.4$_{-0.6}^{+0.7}$     &  199$_{-38}^{+45}$     \\
J1913+0618  & 8.9$_{-1.2}^{+1.1}$       & 5.8$_{-0.9}^{+0.9}$     &  245$_{-49}^{+47}$     \\
J1921+1929  & 11.5$_{-0.6}^{+0.6}$      & 2.4$_{-0.4}^{+0.4}$     &  132$_{-21}^{+21}$     \\
J1928+1246  & 4.4$_{-1.3}^{+1.3}$       & 6.0$_{-0.9}^{+0.9}$     &  126$_{-40}^{+42}$      \\
J1930+2441  & 9.7$_{-1.7}^{+1.5}$       & 3.2$_{-0.5}^{+0.5}$     &  153$_{-34}^{+31}$    \\
J1932+1756  & 15.3$_{-3.7}^{+4.1}$      & 2.1$_{-0.3}^{+0.3}$     &  153$_{-44}^{+47}$    \\
J1937+1658  & 5.0$_{-0.6}^{+0.6}$       & 3.3$_{-0.5}^{+0.5}$     &  77$_{-14}^{+14}$      \\
J2010+3051  & 10.1$_{-0.5}^{+0.5}$      & 5.1$_{-0.8}^{+0.8}$     &  244$_{-38}^{+38}$    \\
\hline
  \end{tabular}
  \label{tab:pm}
\end{table}
%\end{document}

%% file: table_pdot.tex
%\begin{document}
\begin{table}[h]
  \centering
  \caption{Contributions to the observed period derivatives $\dot{P}_{obs}$. Numbers in parentheses are 1-$\sigma$ uncertainties in the last digit quoted. Uncertainties on $\dot{P}_{Gal}$, $\dot{P}_{shk}$ and $\dot{P}_{int}$ are calculated using the Monte Carlo method described in Section \ref{fig:ppdot}.} 
  \setlength{\tabcolsep}{1.6mm}
  \begin{tabular}{c c c c c}
\hline
PSR & $\dot{P}_{\rm obs}$ & $\dot{P}_{\rm gal}$ & $\dot{P}_{\rm shk}$  &$\dot{P}_{\rm int}$ \\
    & (10$^{-20}$)    & (10$^{-20}$)    & (10$^{-20}$)     & (10$^{-20}$) \\                 
\hline \hline
J1906+0454    & 0.298(1)    & $-$0.02$_{-0.02}^{+0.007}$  & 0.25$_{-0.04}^{+0.03}$     & 0.07$_{-0.02}^{+0.04}$ \\
J1913+0618    & 0.961(2)    & $-$0.3$_{-0.08}^{+0.07}$    & 0.58$_{-0.13}^{+0.19}$      & 0.64$_{-0.15}^{+0.13}$  \\
J1921+1929    & 3.8196(5)   & $-$0.04$_{-0.01}^{+0.008}$  & 0.21$_{-0.03}^{+0.04}$     & 3.65$_{-0.04}^{+0.03}$ \\
J1928+1246    & 0.168(3)    & $-$0.18$_{-0.03}^{+0.04}$   & 0.11$_{-0.03}^{+0.08}$     & 0.24$_{-0.07}^{+0.06}$ \\
J1930+2441    & 0.866(5)    & $-$0.16$_{-0.04}^{+0.03}$   & 0.45 $_{-0.11}^{+0.16}$     & 0.58$_{-0.15}^{+0.11}$  \\
J1932+1756    & 43.22(5)    & $-$0.51$_{-0.14}^{+0.10}$   & 6.0  $_{-1.7}^{+4.7}$      & 37.7$_{-4.6}^{+1.7}$   \\
J1937+1658    & 1.3203(4)   & $-$0.1$_{-0.03}^{+0.02}$    & 0.08$_{-0.02}^{+0.03}$     & 1.34 $_{-0.02}^{+0.03}$ \\
J2010+3051    & 0.487(2)    & $-$0.23$_{-0.02}^{+0.04}$   & 0.60$_{-0.09}^{+0.05}$      & 0.12$_{-0.03}^{+0.06}$  \\
\hline
  \end{tabular}
  \label{tab:pdot}
\end{table}
%\end{document}

%% file: table_heights2.tex
\begin{table}[H]
  \centering
  \caption{Statistical height distributions for the five binary-MSP classes, summarizing results shown in Figure \ref{fig:zz_mf}. \textit{N}, $|z_{mean}|$ and $\sigma$ are the number of systems, the mean absolute height and the corresponding standard deviation in $| z |$ for each binary class, respectively. } 
  \begin{tabular}{l p{1.2cm} c p{1.2cm} c c}
  \hline
  & & \multicolumn{2}{l}{\,\,\,\,\,\,\,YMW16} & \multicolumn{2}{c}{NE2001} \\
  %              &               &     YMW16      &          &  NE2001        &         \\
Binary class  &    \textit{N} & $| z_{mean} |$ & $\sigma$ & $| z_{mean} |$ & $\sigma$ \\
              &               & (kpc)          & (kpc)    & (kpc)          & (kpc)  \\  
\hline \hline
UL          &  31       &   0.72    &   0.54    &   0.51    &  0.23    \\
He WD       &  118      &   0.42    &   0.46    &   0.34    &  0.24     \\
CO/ONeMg WD &  34       &   0.25    &   0.30    &   0.21    &  0.18     \\
NS          &  14       &   0.33    &   0.38    &   0.27    &  0.27     \\
MS          &  15       &   0.67    &   0.56    &   0.39    &  0.24      \\
\hline
  \end{tabular}
  \label{tab:height2}
\end{table}

%% file: PALFA-MSP-Timing-v2.bbl
\begin{thebibliography}{}
\expandafter\ifx\csname natexlab\endcsname\relax\def\natexlab#1{#1}\fi

\bibitem[{{Abbott} {et~al.}(2009){Abbott}, {Abbott}, {Adhikari}, {Ajith},
  {Allen}, {Allen}, {Amin}, {Anderson}, {Anderson}, \& {Arain}}]{aaa+09}
{Abbott}, B.~P., {Abbott}, R., {Adhikari}, R., {et~al.} 2009, Reports on
  Progress in Physics, 72, 076901

\bibitem[{{Abdo} {et~al.}(2009){Abdo}, {Ackermann}, {Ajello}, {Anderson},
  {Atwood}, {Axelsson}, {Baldini}, {Ballet}, {Barbiellini}, {Baring},
  {Bastieri}, {Baughman}, {Bechtol}, {Bellazzini}, {Berenji}, {Bignami},
  {Blandford}, {Bloom}, {Bonamente}, {Borgland}, {Bregeon}, {Brez}, {Brigida},
  {Bruel}, {Burnett}, {Caliandro}, {Cameron}, {Caraveo}, {Casandjian},
  {Cecchi}, {{\c{C}}elik}, {Chekhtman}, {Cheung}, {Chiang}, {Ciprini}, {Claus},
  {Cohen-Tanugi}, {Conrad}, {Cutini}, {Dermer}, {de Angelis}, {de Luca}, {de
  Palma}, {Digel}, {Dormody}, {do Couto e Silva}, {Drell}, {Dubois}, {Dumora},
  {Farnier}, {Favuzzi}, {Fegan}, {Fukazawa}, {Funk}, {Fusco}, {Gargano},
  {Gasparrini}, {Gehrels}, {Germani}, {Giebels}, {Giglietto}, {Giommi},
  {Giordano}, {Glanzman}, {Godfrey}, {Grenier}, {Grondin}, {Grove},
  {Guillemot}, {Guiriec}, {Gwon}, {Hanabata}, {Harding}, {Hayashida}, {Hays},
  {Hughes}, {J{\'o}hannesson}, {Johnson}, {Johnson}, {Johnson}, {Kamae},
  {Katagiri}, {Kataoka}, {Kawai}, {Kerr}, {Kn{\"o}dlseder}, {Kocian}, {Kuss},
  {Lande}, {Latronico}, {Lemoine-Goumard}, {Longo}, {Loparco}, {Lott},
  {Lovellette}, {Lubrano}, {Madejski}, {Makeev}, {Marelli}, {Mazziotta},
  {McConville}, {McEnery}, {Meurer}, {Michelson}, {Mitthumsiri}, {Mizuno},
  {Monte}, {Monzani}, {Morselli}, {Moskalenko}, {Murgia}, {Nolan}, {Norris},
  {Nuss}, {Ohsugi}, {Omodei}, {Orlando}, {Ormes}, {Paneque}, {Parent},
  {Pelassa}, {Pepe}, {Pesce-Rollins}, {Pierbattista}, {Piron}, {Porter},
  {Primack}, {Rain{\`o}}, {Rando}, {Ray}, {Razzano}, {Rea}, {Reimer}, {Reimer},
  {Reposeur}, {Ritz}, {Rochester}, {Rodriguez}, {Romani}, {Ryde},
  {Sadrozinski}, {Sanchez}, {Sander}, {Parkinson}, {Scargle}, {Sgr{\`o}},
  {Siskind}, {Smith}, {Smith}, {Spandre}, {Spinelli}, {Starck}, {Strickman},
  {Suson}, {Tajima}, {Takahashi}, {Takahashi}, {Tanaka}, {Thayer}, {Thompson},
  {Tibaldo}, {Tibolla}, {Torres}, {Tosti}, {Tramacere}, {Uchiyama}, {Usher},
  {Van Etten}, {Vasileiou}, {Vilchez}, {Vitale}, {Waite}, {Wang}, {Watters},
  {Winer}, {Wolff}, {Wood}, {Ylinen}, {Ziegler}, \& {Fermi LAT
  Collaboration}}]{fermi09}
{Abdo}, A.~A., {Ackermann}, M., {Ajello}, M., {et~al.} 2009, Science, 325, 840

\bibitem[{Abdo {et~al.}(2010)Abdo, Ackermann, Ajello, Atwood, Axelsson,
  Baldini, Ballet, Barbiellini, Baring, Bastieri, Baughman, Bechtol,
  Bellazzini, Berenji, Blandford, Bloom, Bonamente, Borgland, Bregeon, Brez,
  Brigida, Bruel, Burnett, Buson, Caliandro, Cameron, Camilo, Caraveo,
  Casandjian, Cecchi, Çelik, Charles, Chekhtman, Cheung, Chiang, Ciprini,
  Claus, Cognard, Cohen-Tanugi, Cominsky, Conrad, Corbet, Cutini, den Hartog,
  Dermer, de~Angelis, de~Luca, de~Palma, Digel, Dormody, do~Couto~e Silva,
  Drell, Dubois, Dumora, Espinoza, Farnier, Favuzzi, Fegan, Ferrara, Focke,
  Fortin, Frailis, Freire, Fukazawa, Funk, Fusco, Gargano, Gasparrini, Gehrels,
  Germani, Giavitto, Giebels, Giglietto, Giommi, Giordano, Glanzman, Godfrey,
  Gotthelf, Grenier, Grondin, Grove, Guillemot, Guiriec, Gwon, Hanabata,
  Harding, Hayashida, Hays, Hughes, Jackson, Jóhannesson, Johnson, Johnson,
  Johnson, Johnson, Johnston, Kamae, Kanbach, Kaspi, Katagiri, Kataoka, Kawai,
  Kerr, Knödlseder, Kocian, Kramer, Kuss, Lande, Latronico, Lemoine-Goumard,
  Livingstone, Longo, Loparco, Lott, Lovellette, Lubrano, Lyne, Madejski,
  Makeev, Manchester, Marelli, Mazziotta, McConville, McEnery, McGlynn, Meurer,
  Michelson, Mineo, Mitthumsiri, Mizuno, Moiseev, Monte, Monzani, Morselli,
  Moskalenko, Murgia, Nakamori, Nolan, Norris, Noutsos, Nuss, Ohsugi, Omodei,
  Orlando, Ormes, Ozaki, Paneque, Panetta, Parent, Pelassa, Pepe,
  Pesce-Rollins, Piron, Porter, Rainò, Rando, Ransom, Ray, Razzano, Rea,
  Reimer, Reimer, Reposeur, Ritz, Rodriguez, Romani, Roth, Ryde, Sadrozinski,
  Sanchez, Sander, Parkinson, Scargle, Schalk, Sellerholm, Sgrò, Siskind,
  Smith, Smith, Spandre, Spinelli, Stappers, Starck, Striani, Strickman,
  Strong, Suson, Tajima, Takahashi, Takahashi, Tanaka, Thayer, Thayer,
  Theureau, Thompson, Thorsett, Tibaldo, Tibolla, Torres, Tosti, Tramacere,
  Uchiyama, Usher, Etten, Vasileiou, Venter, Vilchez, Vitale, Waite, Wang,
  Wang, Watters, Weltevrede, Winer, Wood, Ylinen, \& Ziegler}]{fermi10}
Abdo, A.~A., Ackermann, M., Ajello, M., {et~al.} 2010, The Astrophysical
  Journal Supplement Series, 187, 460

\bibitem[{{Abdo} {et~al.}(2013){Abdo}, {Ajello}, {Allafort}, {Baldini},
  {Ballet}, {Barbiellini}, {Baring}, {Bastieri}, {Belfiore}, {Bellazzini},
  {Bhattacharyya}, {Bissaldi}, {Bloom}, {Bonamente}, {Bottacini}, {Brandt},
  {Bregeon}, {Brigida}, {Bruel}, {Buehler}, {Burgay}, {Burnett}, {Busetto},
  {Buson}, {Caliandro}, {Cameron}, {Camilo}, {Caraveo}, {Casandjian}, {Cecchi},
  {{\c{C}}elik}, {Charles}, {Chaty}, {Chaves}, {Chekhtman}, {Chen}, {Chiang},
  {Chiaro}, {Ciprini}, {Claus}, {Cognard}, {Cohen-Tanugi}, {Cominsky},
  {Conrad}, {Cutini}, {D'Ammando}, {de Angelis}, {DeCesar}, {De Luca}, {den
  Hartog}, {de Palma}, {Dermer}, {Desvignes}, {Digel}, {Di Venere}, {Drell},
  {Drlica-Wagner}, {Dubois}, {Dumora}, {Espinoza}, {Falletti}, {Favuzzi},
  {Ferrara}, {Focke}, {Franckowiak}, {Freire}, {Funk}, {Fusco}, {Gargano},
  {Gasparrini}, {Germani}, {Giglietto}, {Giommi}, {Giordano}, {Giroletti},
  {Glanzman}, {Godfrey}, {Gotthelf}, {Grenier}, {Grondin}, {Grove},
  {Guillemot}, {Guiriec}, {Hadasch}, {Hanabata}, {Harding}, {Hayashida},
  {Hays}, {Hessels}, {Hewitt}, {Hill}, {Horan}, {Hou}, {Hughes}, {Jackson},
  {Janssen}, {Jogler}, {J{\'o}hannesson}, {Johnson}, {Johnson}, {Johnson},
  {Johnson}, {Johnston}, {Kamae}, {Kataoka}, {Keith}, {Kerr}, {Kn{\"o}dlseder},
  {Kramer}, {Kuss}, {Lande}, {Larsson}, {Latronico}, {Lemoine- Goumard},
  {Longo}, {Loparco}, {Lovellette}, {Lubrano}, {Lyne}, {Manchester}, {Marelli},
  {Massaro}, {Mayer}, {Mazziotta}, {McEnery}, {McLaughlin}, {Mehault},
  {Michelson}, {Mignani}, {Mitthumsiri}, {Mizuno}, {Moiseev}, {Monzani},
  {Morselli}, {Moskalenko}, {Murgia}, {Nakamori}, {Nemmen}, {Nuss}, {Ohno},
  {Ohsugi}, {Orienti}, {Orlando}, {Ormes}, {Paneque}, {Panetta}, {Parent},
  {Perkins}, {Pesce-Rollins}, {Pierbattista}, {Piron}, {Pivato}, {Pletsch},
  {Porter}, {Possenti}, {Rain{\`o}}, {Rando}, {Ransom}, {Ray}, {Razzano},
  {Rea}, {Reimer}, {Reimer}, {Renault}, {Reposeur}, {Ritz}, {Romani}, {Roth},
  {Rousseau}, {Roy}, {Ruan}, {Sartori}, {Saz Parkinson}, {Scargle}, {Schulz},
  {Sgr{\`o}}, {Shannon}, {Siskind}, {Smith}, {Spandre}, {Spinelli}, {Stappers},
  {Strong}, {Suson}, {Takahashi}, {Thayer}, {Thayer}, {Theureau}, {Thompson},
  {Thorsett}, {Tibaldo}, {Tibolla}, {Tinivella}, {Torres}, {Tosti}, {Troja},
  {Uchiyama}, {Usher}, {Vandenbroucke}, {Vasileiou}, {Venter}, {Vianello},
  {Vitale}, {Wang}, {Weltevrede}, {Winer}, {Wolff}, {Wood}, {Wood}, {Wood}, \&
  {Yang}}]{fermi13}
{Abdo}, A.~A., {Ajello}, M., {Allafort}, A., {et~al.} 2013, The Astrophysical
  Journal Supplement Series, 208, 17

\bibitem[{{Acero} {et~al.}(2015{\natexlab{a}}){Acero}, {Ackermann}, {Ajello},
  {Albert}, {Atwood}, {Axelsson}, {Baldini}, {Ballet}, {Barbiellini},
  {Bastieri}, {Belfiore}, {Bellazzini}, {Bissaldi}, {Blandford}, {Bloom},
  {Bogart}, {Bonino}, {Bottacini}, {Bregeon}, {Britto}, {Bruel}, {Buehler},
  {Burnett}, {Buson}, {Caliandro}, {Cameron}, {Caputo}, {Caragiulo}, {Caraveo},
  {Casandjian}, {Cavazzuti}, {Charles}, {Chaves}, {Chekhtman}, {Cheung},
  {Chiang}, {Chiaro}, {Ciprini}, {Claus}, {Cohen-Tanugi}, {Cominsky}, {Conrad},
  {Cutini}, {D'Ammando}, {de Angelis}, {DeKlotz}, {de Palma}, {Desiante},
  {Digel}, {Di Venere}, {Drell}, {Dubois}, {Dumora}, {Favuzzi}, {Fegan},
  {Ferrara}, {Finke}, {Franckowiak}, {Fukazawa}, {Funk}, {Fusco}, {Gargano},
  {Gasparrini}, {Giebels}, {Giglietto}, {Giommi}, {Giordano}, {Giroletti},
  {Glanzman}, {Godfrey}, {Grenier}, {Grondin}, {Grove}, {Guillemot}, {Guiriec},
  {Hadasch}, {Harding}, {Hays}, {Hewitt}, {Hill}, {Horan}, {Iafrate}, {Jogler},
  {J{\'o}hannesson}, {Johnson}, {Johnson}, {Johnson}, {Johnson}, {Kamae},
  {Kataoka}, {Katsuta}, {Kuss}, {La Mura}, {Landriu}, {Larsson}, {Latronico},
  {Lemoine-Goumard}, {Li}, {Li}, {Longo}, {Loparco}, {Lott}, {Lovellette},
  {Lubrano}, {Madejski}, {Massaro}, {Mayer}, {Mazziotta}, {McEnery},
  {Michelson}, {Mirabal}, {Mizuno}, {Moiseev}, {Mongelli}, {Monzani},
  {Morselli}, {Moskalenko}, {Murgia}, {Nuss}, {Ohno}, {Ohsugi}, {Omodei},
  {Orienti}, {Orlando}, {Ormes}, {Paneque}, {Panetta}, {Perkins},
  {Pesce-Rollins}, {Piron}, {Pivato}, {Porter}, {Racusin}, {Rando}, {Razzano},
  {Razzaque}, {Reimer}, {Reimer}, {Reposeur}, {Rochester}, {Romani},
  {Salvetti}, {S{\'a}nchez-Conde}, {Saz Parkinson}, {Schulz}, {Siskind},
  {Smith}, {Spada}, {Spandre}, {Spinelli}, {Stephens}, {Strong}, {Suson},
  {Takahashi}, {Takahashi}, {Tanaka}, {Thayer}, {Thayer}, {Thompson},
  {Tibaldo}, {Tibolla}, {Torres}, {Torresi}, {Tosti}, {Troja}, {Van Klaveren},
  {Vianello}, {Winer}, {Wood}, {Wood}, {Zimmer}, \& {Fermi-LAT
  Collaboration}}]{fermi4yr}
{Acero}, F., {Ackermann}, M., {Ajello}, M., {et~al.} 2015{\natexlab{a}}, The
  Astrophysical Journal Supplement Series, 218, 23

\bibitem[{{Acero} {et~al.}(2015{\natexlab{b}}){Acero}, {Ackermann}, {Ajello},
  {Albert}, {Atwood}, {Axelsson}, {Baldini}, {Ballet}, {Barbiellini},
  {Bastieri}, {Belfiore}, {Bellazzini}, {Bissaldi}, {Blandford}, {Bloom},
  {Bogart}, {Bonino}, {Bottacini}, {Bregeon}, {Britto}, {Bruel}, {Buehler},
  {Burnett}, {Buson}, {Caliandro}, {Cameron}, {Caputo}, {Caragiulo}, {Caraveo},
  {Casandjian}, {Cavazzuti}, {Charles}, {Chaves}, {Chekhtman}, {Cheung},
  {Chiang}, {Chiaro}, {Ciprini}, {Claus}, {Cohen-Tanugi}, {Cominsky}, {Conrad},
  {Cutini}, {D'Ammando}, {de Angelis}, {DeKlotz}, {de Palma}, {Desiante},
  {Digel}, {Di Venere}, {Drell}, {Dubois}, {Dumora}, {Favuzzi}, {Fegan},
  {Ferrara}, {Finke}, {Franckowiak}, {Fukazawa}, {Funk}, {Fusco}, {Gargano},
  {Gasparrini}, {Giebels}, {Giglietto}, {Giommi}, {Giordano}, {Giroletti},
  {Glanzman}, {Godfrey}, {Grenier}, {Grondin}, {Grove}, {Guillemot}, {Guiriec},
  {Hadasch}, {Harding}, {Hays}, {Hewitt}, {Hill}, {Horan}, {Iafrate}, {Jogler},
  {J{\'o}hannesson}, {Johnson}, {Johnson}, {Johnson}, {Johnson}, {Kamae},
  {Kataoka}, {Katsuta}, {Kuss}, {La Mura}, {Landriu}, {Larsson}, {Latronico},
  {Lemoine-Goumard}, {Li}, {Li}, {Longo}, {Loparco}, {Lott}, {Lovellette},
  {Lubrano}, {Madejski}, {Massaro}, {Mayer}, {Mazziotta}, {McEnery},
  {Michelson}, {Mirabal}, {Mizuno}, {Moiseev}, {Mongelli}, {Monzani},
  {Morselli}, {Moskalenko}, {Murgia}, {Nuss}, {Ohno}, {Ohsugi}, {Omodei},
  {Orienti}, {Orlando}, {Ormes}, {Paneque}, {Panetta}, {Perkins},
  {Pesce-Rollins}, {Piron}, {Pivato}, {Porter}, {Racusin}, {Rando}, {Razzano},
  {Razzaque}, {Reimer}, {Reimer}, {Reposeur}, {Rochester}, {Romani},
  {Salvetti}, {S{\'a}nchez-Conde}, {Saz Parkinson}, {Schulz}, {Siskind},
  {Smith}, {Spada}, {Spandre}, {Spinelli}, {Stephens}, {Strong}, {Suson},
  {Takahashi}, {Takahashi}, {Tanaka}, {Thayer}, {Thayer}, {Thompson},
  {Tibaldo}, {Tibolla}, {Torres}, {Torresi}, {Tosti}, {Troja}, {Van Klaveren},
  {Vianello}, {Winer}, {Wood}, {Wood}, {Zimmer}, \& {Fermi-LAT
  Collaboration}}]{fermi15}
---. 2015{\natexlab{b}}, The Astrophysical Journal Supplement Series, 218, 23

\bibitem[{Andrews {et~al.}(2015)Andrews, Farr, Kalogera, \& Willems}]{afk+15}
Andrews, J.~J., Farr, W.~M., Kalogera, V., \& Willems, B. 2015, The
  Astrophysical Journal, 801, 32

\bibitem[{{Antoniadis} {et~al.}(2016){Antoniadis}, {Tauris}, {Ozel}, {Barr},
  {Champion}, \& {Freire}}]{ato+16}
{Antoniadis}, J., {Tauris}, T.~M., {Ozel}, F., {et~al.} 2016, arXiv e-prints,
  arXiv:1605.01665

\bibitem[{{Arzoumanian} {et~al.}(2018){Arzoumanian}, {Brazier},
  {Burke-Spolaor}, {Chamberlin}, {Chatterjee}, {Christy}, {Cordes}, {Cornish},
  {Crawford}, {Thankful Cromartie}, {Crowter}, {DeCesar}, {Demorest}, {Dolch},
  {Ellis}, {Ferdman}, {Ferrara}, {Fonseca}, {Garver-Daniels}, {Gentile},
  {Halmrast}, {Huerta}, {Jenet}, {Jessup}, {Jones}, {Jones}, {Kaplan}, {Lam},
  {Lazio}, {Levin}, {Lommen}, {Lorimer}, {Luo}, {Lynch}, {Madison}, {Matthews},
  {McLaughlin}, {McWilliams}, {Mingarelli}, {Ng}, {Nice}, {Pennucci}, {Ransom},
  {Ray}, {Siemens}, {Simon}, {Spiewak}, {Stairs}, {Stinebring}, {Stovall},
  {Swiggum}, {Taylor}, {Vallisneri}, {van Haasteren}, {Vigeland}, {Zhu}, \&
  {NANOGrav Collaboration}}]{abb+18}
{Arzoumanian}, Z., {Brazier}, A., {Burke-Spolaor}, S., {et~al.} 2018, \apjs,
  235, 37

\bibitem[{Beniamini \& Piran(2016)}]{bp16}
Beniamini, P., \& Piran, T. 2016, Monthly Notices of the Royal Astronomical
  Society, 456, 4089

\bibitem[{{Boller} {et~al.}(2016){Boller}, {Freyberg}, {Tr{\"u}mper}, {Haberl},
  {Voges}, \& {Nandra}}]{bft+16}
{Boller}, T., {Freyberg}, M.~J., {Tr{\"u}mper}, J., {et~al.} 2016, \aap, 588,
  A103

\bibitem[{{Burgay} {et~al.}(2003){Burgay}, {D'Amico}, {Possenti}, {Manchester},
  {Lyne}, {Joshi}, {McLaughlin}, {Kramer}, {Sarkissian}, \& {Camilo}}]{bap+03}
{Burgay}, M., {D'Amico}, N., {Possenti}, A., {et~al.} 2003, \nat, 426, 531

\bibitem[{{Cameron} {et~al.}(2018){Cameron}, {Champion}, {Kramer}, {Bailes},
  {Barr}, {Bassa}, {Bhandari}, {Bhat}, {Burgay}, \& {Burke-Spolaor}}]{cck+18}
{Cameron}, A.~D., {Champion}, D.~J., {Kramer}, M., {et~al.} 2018, \mnras, 475,
  L57

\bibitem[{{Camilo} {et~al.}(2001){Camilo}, {Lyne}, {Manchester}, {Bell},
  {Stairs}, {D'Amico}, {Kaspi}, {Possenti}, {Crawford}, \& {McKay}}]{clm+01}
{Camilo}, F., {Lyne}, A.~G., {Manchester}, R.~N., {et~al.} 2001, \apjl, 548,
  L187

\bibitem[{{Champion} {et~al.}(2008){Champion}, {Ransom}, {Lazarus}, {Camilo},
  {Bassa}, {Kaspi}, {Nice}, {Freire}, {Stairs}, {van Leeuwen}, {Stappers},
  {Cordes}, {Hessels}, {Lorimer}, {Arzoumanian}, {Backer}, {Bhat},
  {Chatterjee}, {Cognard}, {Deneva}, {Faucher-Gigu{\`e}re}, {Gaensler}, {Han},
  {Jenet}, {Kasian}, {Kondratiev}, {Kramer}, {Lazio}, {McLaughlin},
  {Venkataraman}, \& {Vlemmings}}]{crl+08}
{Champion}, D.~J., {Ransom}, S.~M., {Lazarus}, P., {et~al.} 2008, Science, 320,
  1309

\bibitem[{Coifman \& Donoho(1995)}]{cd95}
Coifman, R.~R., \& Donoho, D.~L. 1995, Translation-Invariant De-Noising, ed.
  A.~Antoniadis \& G.~Oppenheim (New York, NY: Springer New York), 125--150

\bibitem[{{Cordes} \& {Chernoff}(1998)}]{cc98}
{Cordes}, J.~M., \& {Chernoff}, D.~F. 1998, \apj, 505, 315

\bibitem[{{Cordes} \& {Lazio}(2002)}]{cl02}
{Cordes}, J.~M., \& {Lazio}, T.~J.~W. 2002, ArXiv Astrophysics e-prints,
  astro-ph/0207156

\bibitem[{{Cordes} {et~al.}(2006){Cordes}, {Freire}, {Lorimer}, {Camilo},
  {Champion}, {Nice}, {Ramachandran}, {Hessels}, {Vlemmings}, {van Leeuwen},
  {Ransom}, {Bhat}, {Arzoumanian}, {McLaughlin}, {Kaspi}, {Kasian}, {Deneva},
  {Reid}, {Chatterjee}, {Han}, {Backer}, {Stairs}, {Deshpande}, \&
  {Faucher-Gigu{\`e}re}}]{cfl+06}
{Cordes}, J.~M., {Freire}, P.~C.~C., {Lorimer}, D.~R., {et~al.} 2006, \apj,
  637, 446

\bibitem[{{Crawford} {et~al.}(2012){Crawford}, {Stovall}, {Lyne}, {Stappers},
  {Nice}, {Stairs}, {Lazarus}, {Hessels}, {Freire}, {Allen}, {Bhat},
  {Bogdanov}, {Brazier}, {Camilo}, {Champion}, {Chatterjee}, {Cognard},
  {Cordes}, {Deneva}, {Desvignes}, {Jenet}, {Kaspi}, {Knispel}, {Kramer}, {van
  Leeuwen}, {Lorimer}, {Lynch}, {McLaughlin}, {Ransom}, {Scholz}, {Siemens}, \&
  {Venkataraman}}]{csl+12}
{Crawford}, F., {Stovall}, K., {Lyne}, A.~G., {et~al.} 2012, \apj, 757, 90

\bibitem[{{Cromartie} {et~al.}(2016){Cromartie}, {Camilo}, {Kerr}, {Deneva},
  {Ransom}, {Ray}, {Ferrara}, {Michelson}, \& {Wood}}]{cck+16}
{Cromartie}, H.~T., {Camilo}, F., {Kerr}, M., {et~al.} 2016, \apj, 819, 34

\bibitem[{Damour \& Deruelle(1986)}]{dd86}
Damour, T., \& Deruelle, N. 1986, Annales de l'I.H.P. Physique th\'eorique, 44,
  263

\bibitem[{{Damour} \& {Taylor}(1991)}]{dt91}
{Damour}, T., \& {Taylor}, J.~H. 1991, \apj, 366, 501

\bibitem[{{de Jager} \& {B{\"u}sching}(2010)}]{db10}
{de Jager}, O.~C., \& {B{\"u}sching}, I. 2010, \aap, 517, L9

\bibitem[{{Deller} {et~al.}(2019){Deller}, {Goss}, {Brisken}, {Chatterjee},
  {Cordes}, {Janssen}, {Kovalev}, {Lazio}, {Petrov}, {Stappers}, \&
  {Lyne}}]{dgb+19}
{Deller}, A.~T., {Goss}, W.~M., {Brisken}, W.~F., {et~al.} 2019, \apj, 875, 100

\bibitem[{{Demorest} {et~al.}(2013){Demorest}, {Ferdman}, {Gonzalez}, {Nice},
  {Ransom}, {Stairs}, {Arzoumanian}, {Brazier}, {Burke-Spolaor}, {Chamberlin},
  {Cordes}, {Ellis}, {Finn}, {Freire}, {Giampanis}, {Jenet}, {Kaspi}, {Lazio},
  {Lommen}, {McLaughlin}, {Palliyaguru}, {Perrodin}, {Shannon}, {Siemens},
  {Stinebring}, {Swiggum}, \& {Zhu}}]{dfg+13}
{Demorest}, P.~B., {Ferdman}, R.~D., {Gonzalez}, M.~E., {et~al.} 2013, \apj,
  762, 94

\bibitem[{{Deneva} {et~al.}(2012){Deneva}, {Freire}, {Cordes}, {Lyne},
  {Ransom}, {Cognard}, {Camilo}, {Nice}, {Stairs}, {Allen}, {Bhat}, {Bogdanov},
  {Brazier}, {Champion}, {Chatterjee}, {Crawford}, {Desvignes}, {Hessels},
  {Jenet}, {Kaspi}, {Knispel}, {Kramer}, {Lazarus}, {van Leeuwen}, {Lorimer},
  {Lynch}, {McLaughlin}, {Scholz}, {Siemens}, {Stappers}, {Stovall}, \&
  {Venkataraman}}]{dfc+12}
{Deneva}, J.~S., {Freire}, P.~C.~C., {Cordes}, J.~M., {et~al.} 2012, \apj, 757,
  89

\bibitem[{{Dewi} {et~al.}(2005){Dewi}, {Podsiadlowski}, \& {Pols}}]{dpp05}
{Dewi}, J.~D.~M., {Podsiadlowski}, P., \& {Pols}, O.~R. 2005, \mnras, 363, L71

\bibitem[{{Eichler} \& {Levinson}(1988)}]{el88}
{Eichler}, D., \& {Levinson}, A. 1988, \apjl, 335, L67

\bibitem[{{Fonseca} {et~al.}(2014){Fonseca}, {Stairs}, \& {Thorsett}}]{fst+14}
{Fonseca}, E., {Stairs}, I.~H., \& {Thorsett}, S.~E. 2014, \apj, 787, 82

\bibitem[{{Foster} \& {Backer}(1990)}]{fb90}
{Foster}, R.~S., \& {Backer}, D.~C. 1990, \apj, 361, 300

\bibitem[{{Freire}(2005)}]{f05}
{Freire}, P.~C.~C. 2005, in Astronomical Society of the Pacific Conference
  Series, Vol. 328, Binary Radio Pulsars, ed. F.~A. {Rasio} \& I.~H. {Stairs},
  405

\bibitem[{{Freire} \& {Ridolfi}(2018)}]{fr18}
{Freire}, P. C.~C., \& {Ridolfi}, A. 2018, \mnras, 476, 4794

\bibitem[{{Freire} {et~al.}(2011){Freire}, {Bassa}, {Wex}, {Stairs},
  {Champion}, {Ransom}, {Lazarus}, {Kaspi}, {Hessels}, {Kramer}, {Cordes},
  {Verbiest}, {Podsiadlowski}, {Nice}, {Deneva}, {Lorimer}, {Stappers},
  {McLaughlin}, \& {Camilo}}]{fbw+11}
{Freire}, P.~C.~C., {Bassa}, C.~G., {Wex}, N., {et~al.} 2011, \mnras, 412, 2763

\bibitem[{{Fruchter} {et~al.}(1990){Fruchter}, {Berman}, {Bower}, {Convery},
  {Goss}, {Hankins}, {Klein}, {Nice}, {Ryba}, {Stinebring}, {Taylor},
  {Thorsett}, \& {Weisberg}}]{fbb+90}
{Fruchter}, A.~S., {Berman}, G., {Bower}, G., {et~al.} 1990, \apj, 351, 642

\bibitem[{{Gaia Collaboration} {et~al.}(2017){Gaia Collaboration},
  {Clementini}, {Eyer}, {Ripepi}, {Marconi}, {Muraveva}, {Garofalo}, {Sarro},
  {Palmer}, {Luri}, \& et~al.}]{gaia}
{Gaia Collaboration}, {Clementini}, G., {Eyer}, L., {et~al.} 2017, \aap, 605,
  A79

\bibitem[{{Gentile} {et~al.}(2018){Gentile}, {McLaughlin}, {Demorest},
  {Stairs}, {Arzoumanian}, {Crowter}, {Dolch}, {DeCesar}, {Ellis}, {Ferdman},
  {Ferrara}, {Fonseca}, {Gonzalez}, {Jones}, {Jones}, {Lam}, {Levin},
  {Lorimer}, {Lynch}, {Ng}, {Nice}, {Pennucci}, {Ransom}, {Ray}, {Spiewak},
  {Stovall}, {Swiggum}, \& {Zhu}}]{gmd+18}
{Gentile}, P.~A., {McLaughlin}, M.~A., {Demorest}, P.~B., {et~al.} 2018, \apj,
  862, 47

\bibitem[{{Guillemot} {et~al.}(2019){Guillemot}, {Octau}, {Cognard},
  {Desvignes}, {Freire}, {Smith}, {Theureau}, \& {Burnett}}]{goc+19}
{Guillemot}, L., {Octau}, F., {Cognard}, I., {et~al.} 2019, arXiv e-prints,
  arXiv:1907.09778

\bibitem[{{Hellings} \& {Downs}(1983)}]{hd83}
{Hellings}, R.~W., \& {Downs}, G.~S. 1983, \apj, 265, L39

\bibitem[{{Hessels} {et~al.}(2011){Hessels}, {Roberts}, {McLaughlin}, {Ray},
  {Bangale}, {Ransom}, {Kerr}, {Camilo}, \& {Decesar}}]{hrm+11}
{Hessels}, J.~W.~T., {Roberts}, M.~S.~E., {McLaughlin}, M.~A., {et~al.} 2011,
  in American Institute of Physics Conference Series, Vol. 1357, American
  Institute of Physics Conference Series, ed. M.~{Burgay}, N.~{D'Amico},
  P.~{Esposito}, A.~{Pellizzoni}, \& A.~{Possenti}, 40--43

\bibitem[{{Hobbs} {et~al.}(2010){Hobbs}, {Archibald}, {Arzoumanian}, {Backer},
  {Bailes}, {Bhat}, {Burgay}, {Burke-Spolaor}, {Champion}, {Cognard}, {Coles},
  {Cordes}, {Demorest}, {Desvignes}, {Ferdman}, {Finn}, {Freire}, {Gonzalez},
  {Hessels}, {Hotan}, {Janssen}, {Jenet}, {Jessner}, {Jordan}, {Kaspi},
  {Kramer}, {Kondratiev}, {Lazio}, {Lazaridis}, {Lee}, {Levin}, {Lommen},
  {Lorimer}, {Lynch}, {Lyne}, {Manchester}, {McLaughlin}, {Nice}, {Oslowski},
  {Pilia}, {Possenti}, {Purver}, {Ransom}, {Reynolds}, {Sanidas}, {Sarkissian},
  {Sesana}, {Shannon}, {Siemens}, {Stairs}, {Stappers}, {Stinebring},
  {Theureau}, {van Haasteren}, {van Straten}, {Verbiest}, {Yardley}, \&
  {You}}]{haa+10}
{Hobbs}, G., {Archibald}, A., {Arzoumanian}, Z., {et~al.} 2010, Classical and
  Quantum Gravity, 27, 084013

\bibitem[{{Hui} {et~al.}(2018){Hui}, {Wu}, {Han}, {Kong}, \& {Tam}}]{hwh+18}
{Hui}, C.~Y., {Wu}, K., {Han}, Q., {Kong}, A.~K.~H., \& {Tam}, P.~H.~T. 2018,
  \apj, 864, 30

\bibitem[{{Jaffe} \& {Backer}(2003)}]{jb03}
{Jaffe}, A.~H., \& {Backer}, D.~C. 2003, \apj, 583, 616

\bibitem[{{Khargharia} {et~al.}(2012){Khargharia}, {Stocke}, {Froning},
  {Gopakumar}, \& {Joshi}}]{ksf+12}
{Khargharia}, J., {Stocke}, J.~T., {Froning}, C.~S., {Gopakumar}, A., \&
  {Joshi}, B.~C. 2012, \apj, 744, 183

\bibitem[{{Kiel} \& {Hurley}(2009)}]{kh09}
{Kiel}, P.~D., \& {Hurley}, J.~R. 2009, \mnras, 395, 2326

\bibitem[{Kiziltan {et~al.}(2013)Kiziltan, Kottas, Yoreo, \& Thorsett}]{kky+13}
Kiziltan, B., Kottas, A., Yoreo, M.~D., \& Thorsett, S.~E. 2013, The
  Astrophysical Journal, 778, 66

\bibitem[{{Knispel} {et~al.}(2015){Knispel}, {Lyne}, {Stappers}, {Freire},
  {Lazarus}, {Allen}, {Aulbert}, {Bock}, {Bogdanov}, {Brazier}, {Camilo},
  {Cardoso}, {Chatterjee}, {Cordes}, {Crawford}, {Deneva}, {Eggenstein},
  {Fehrmann}, {Ferdman}, {Hessels}, {Jenet}, {Karako-Argaman}, {Kaspi}, {van
  Leeuwen}, {Lorimer}, {Lynch}, {Machenschalk}, {Madsen}, {McLaughlin},
  {Patel}, {Ransom}, {Scholz}, {Siemens}, {Spitler}, {Stairs}, {Stovall},
  {Swiggum}, {Venkataraman}, {Wharton}, \& {Zhu}}]{kls+15}
{Knispel}, B., {Lyne}, A.~G., {Stappers}, B.~W., {et~al.} 2015, \apj, 806, 140

\bibitem[{{Lange} {et~al.}(2001){Lange}, {Camilo}, {Wex}, {Kramer}, {Backer},
  {Lyne}, \& {Doroshenko}}]{lcw+01}
{Lange}, C., {Camilo}, F., {Wex}, N., {et~al.} 2001, \mnras, 326, 274

\bibitem[{{Lazaridis} {et~al.}(2009){Lazaridis}, {Wex}, {Jessner}, {Kramer},
  {Stappers}, {Janssen}, {Desvignes}, {Purver}, {Cognard}, {Theureau}, {Lyne},
  {Jordan}, \& {Zensus}}]{lwj+09}
{Lazaridis}, K., {Wex}, N., {Jessner}, A., {et~al.} 2009, \mnras, 400, 805

\bibitem[{{Lazarus} {et~al.}(2015){Lazarus}, {Brazier}, {Hessels},
  {Karako-Argaman}, {Kaspi}, {Lynch}, {Madsen}, {Patel}, {Ransom}, {Scholz},
  {Swiggum}, {Zhu}, {Allen}, {Bogdanov}, {Camilo}, {Cardoso}, {Chatterjee},
  {Cordes}, {Crawford}, {Deneva}, {Ferdman}, {Freire}, {Jenet}, {Knispel},
  {Lee}, {van Leeuwen}, {Lorimer}, {Lyne}, {McLaughlin}, {Siemens}, {Spitler},
  {Stairs}, {Stovall}, \& {Venkataraman}}]{lbh+15}
{Lazarus}, P., {Brazier}, A., {Hessels}, J.~W.~T., {et~al.} 2015, \apj, 812, 81

\bibitem[{{Lazarus} {et~al.}(2016){Lazarus}, {Freire}, {Allen}, {Aulbert},
  {Bock}, {Bogdanov}, {Brazier}, {Camilo}, {Cardoso}, {Chatterjee}, {Cordes},
  {Crawford}, {Deneva}, {Eggenstein}, {Fehrmann}, {Ferdman}, {Hessels},
  {Jenet}, {Karako-Argaman}, {Kaspi}, {Knispel}, {Lynch}, {van Leeuwen},
  {Machenschalk}, {Madsen}, {McLaughlin}, {Patel}, {Ransom}, {Scholz},
  {Seymour}, {Siemens}, {Spitler}, {Stairs}, {Stovall}, {Swiggum},
  {Venkataraman}, \& {Zhu}}]{lfa+16}
{Lazarus}, P., {Freire}, P.~C.~C., {Allen}, B., {et~al.} 2016, \apj, 831, 150

\bibitem[{{Lorimer} {et~al.}(2006){Lorimer}, {Stairs}, {Freire}, {Cordes},
  {Camilo}, {Faulkner}, {Lyne}, {Nice}, {Ransom}, {Arzoumanian}, {Manchester},
  {Champion}, {van Leeuwen}, {Mclaughlin}, {Ramachandran}, {Hessels},
  {Vlemmings}, {Deshpande}, {Bhat}, {Chatterjee}, {Han}, {Gaensler}, {Kasian},
  {Deneva}, {Reid}, {Lazio}, {Kaspi}, {Crawford}, {Lommen}, {Backer}, {Kramer},
  {Stappers}, {Hobbs}, {Possenti}, {D'Amico}, \& {Burgay}}]{lsf+06}
{Lorimer}, D.~R., {Stairs}, I.~H., {Freire}, P.~C., {et~al.} 2006, \apj, 640,
  428

\bibitem[{{Lyne} {et~al.}(2004){Lyne}, {Burgay}, {Kramer}, {Possenti},
  {Manchester}, {Camilo}, {McLaughlin}, {Lorimer}, {D'Amico}, {Joshi},
  {Reynolds}, \& {Freire}}]{lbk+04}
{Lyne}, A.~G., {Burgay}, M., {Kramer}, M., {et~al.} 2004, Science, 303, 1153

\bibitem[{{Manchester}(2017)}]{m17}
{Manchester}, R.~N. 2017, Journal of Astrophysics and Astronomy, 38, 42

\bibitem[{{Manchester} {et~al.}(2005){Manchester}, {Hobbs}, {Teoh}, \&
  {Hobbs}}]{mht+05}
{Manchester}, R.~N., {Hobbs}, G.~B., {Teoh}, A., \& {Hobbs}, M. 2005, VizieR
  Online Data Catalog, 7245

\bibitem[{{Martinez} {et~al.}(2017){Martinez}, {Stovall}, {Freire}, {Deneva},
  {Tauris}, {Ridolfi}, {Wex}, {Jenet}, {McLaughlin}, \& {Bagchi}}]{msf+17}
{Martinez}, J.~G., {Stovall}, K., {Freire}, P.~C.~C., {et~al.} 2017, \apjl,
  851, L29

\bibitem[{{Ng} {et~al.}(2014){Ng}, {Bailes}, {Bates}, {Bhat}, {Burgay},
  {Burke-Spolaor}, {Champion}, {Coster}, {Johnston}, {Keith}, {Kramer},
  {Levin}, {Petroff}, {Possenti}, {Stappers}, {van Straten}, {Thornton},
  {Tiburzi}, {Bassa}, {Freire}, {Guillemot}, {Lyne}, {Tauris}, {Shannon}, \&
  {Wex}}]{nbb+14}
{Ng}, C., {Bailes}, M., {Bates}, S.~D., {et~al.} 2014, \mnras, 439, 1865

\bibitem[{{Nice} {et~al.}(1996){Nice}, {Sayer}, \& {Taylor}}]{nst96}
{Nice}, D.~J., {Sayer}, R.~W., \& {Taylor}, J.~H. 1996, \apjl, 466, L87

\bibitem[{{Nice} \& {Taylor}(1995)}]{nt95}
{Nice}, D.~J., \& {Taylor}, J.~H. 1995, \apj, 441, 429

\bibitem[{{Parent} {et~al.}(2018){Parent}, {Kaspi}, {Ransom}, {Krasteva},
  {Patel}, {Scholz}, {Brazier}, {McLaughlin}, {Boyce}, {Zhu}, {Pleunis},
  {Allen}, {Bogdanov}, {Caballero}, {Camilo}, {Camuccio}, {Chatterjee},
  {Cordes}, {Crawford}, {Deneva}, {Ferdman}, {Freire}, {Hessels}, {Jenet},
  {Knispel}, {Lazarus}, {van Leeuwen}, {Lyne}, {Lynch}, {Seymour}, {Siemens},
  {Stairs}, {Stovall}, \& {Swiggum}}]{pkr+18}
{Parent}, E., {Kaspi}, V.~M., {Ransom}, S.~M., {et~al.} 2018, \apj, 861, 44

\bibitem[{{Patel} {et~al.}(2018){Patel}, {Agarwal}, {Bhardwaj}, {Boyce},
  {Brazier}, {Chatterjee}, {Chawla}, {Kaspi}, {Lorimer}, {McLaughlin},
  {Parent}, {Pleunis}, {Ransom}, {Scholz}, {Wharton}, {Zhu}, {Alam}, {Caballero
  Valdez}, {Camilo}, {Cordes}, {Crawford}, {Deneva}, {Ferdman}, {Freire},
  {Hessels}, {Nguyen}, {Stairs}, {Stovall}, \& {van Leeuwen}}]{pab+18}
{Patel}, C., {Agarwal}, D., {Bhardwaj}, M., {et~al.} 2018, ArXiv e-prints,
  arXiv:1808.03710

\bibitem[{{Phinney}(1992)}]{p92}
{Phinney}, E.~S. 1992, Philosophical Transactions of the Royal Society of
  London Series A, 341, 39

\bibitem[{{Ransom}(2001)}]{r01}
{Ransom}, S.~M. 2001, PhD thesis, Harvard University

\bibitem[{{Ray} {et~al.}(2012){Ray}, {Abdo}, {Parent}, {Bhattacharya},
  {Bhattacharyya}, {Camilo}, {Cognard}, {Theureau}, {Ferrara}, {Harding},
  {Thompson}, {Freire}, {Guillemot}, {Gupta}, {Roy}, {Hessels}, {Johnston},
  {Keith}, {Shannon}, {Kerr}, {Michelson}, {Romani}, {Kramer}, {McLaughlin},
  {Ransom}, {Roberts}, {Saz Parkinson}, {Ziegler}, {Smith}, {Stappers},
  {Weltevrede}, \& {Wood}}]{rap+12}
{Ray}, P.~S., {Abdo}, A.~A., {Parent}, D., {et~al.} 2012, arXiv e-prints,
  arXiv:1205.3089

\bibitem[{{Reid} {et~al.}(2014){Reid}, {Menten}, {Brunthaler}, {Zheng}, {Dame},
  {Xu}, {Wu}, {Zhang}, {Sanna}, {Sato}, {Hachisuka}, {Choi}, {Immer},
  {Moscadelli}, {Rygl}, \& {Bartkiewicz}}]{rmb+14}
{Reid}, M.~J., {Menten}, K.~M., {Brunthaler}, A., {et~al.} 2014, \apj, 783, 130

\bibitem[{{Roberts}(2013)}]{r13}
{Roberts}, M.~S.~E. 2013, in IAU Symposium, Vol. 291, Neutron Stars and
  Pulsars: Challenges and Opportunities after 80 years, ed. J.~{van Leeuwen},
  127--132

\bibitem[{{Saxton} {et~al.}(2008){Saxton}, {Read}, {Esquej}, {Freyberg},
  {Altieri}, \& {Bermejo}}]{sre+08}
{Saxton}, R.~D., {Read}, A.~M., {Esquej}, P., {et~al.} 2008, \aap, 480, 611

\bibitem[{{Scholz} {et~al.}(2015){Scholz}, {Kaspi}, {Lyne}, {Stappers},
  {Bogdanov}, {Cordes}, {Crawford}, {Ferdman}, {Freire}, {Hessels}, {Lorimer},
  {Stairs}, {Allen}, {Brazier}, {Camilo}, {Cardoso}, {Chatterjee}, {Deneva},
  {Jenet}, {Karako-Argaman}, {Knispel}, {Lazarus}, {Lee}, {van Leeuwen},
  {Lynch}, {Madsen}, {McLaughlin}, {Ransom}, {Siemens}, {Spitler}, {Stovall},
  {Swiggum}, {Venkataraman}, \& {Zhu}}]{skl+15}
{Scholz}, P., {Kaspi}, V.~M., {Lyne}, A.~G., {et~al.} 2015, \apj, 800, 123

\bibitem[{{Schwab} {et~al.}(2010){Schwab}, {Podsiadlowski}, \&
  {Rappaport}}]{spr10}
{Schwab}, J., {Podsiadlowski}, P., \& {Rappaport}, S. 2010, \apj, 719, 722

\bibitem[{{Shklovskii}(1970)}]{s70}
{Shklovskii}, I.~S. 1970, \sovast, 13, 562

\bibitem[{{Siemens} {et~al.}(2013){Siemens}, {Ellis}, {Jenet}, \&
  {Romano}}]{sejr13}
{Siemens}, X., {Ellis}, J., {Jenet}, F., \& {Romano}, J.~D. 2013, Classical and
  Quantum Gravity, 30, 224015

\bibitem[{{Skrutskie} {et~al.}(2006){Skrutskie}, {Cutri}, {Stiening},
  {Weinberg}, {Schneider}, {Carpenter}, {Beichman}, {Capps}, {Chester},
  {Elias}, {Huchra}, {Liebert}, {Lonsdale}, {Monet}, {Price}, {Seitzer},
  {Jarrett}, {Kirkpatrick}, {Gizis}, {Howard}, {Evans}, {Fowler}, {Fullmer},
  {Hurt}, {Light}, {Kopan}, {Marsh}, {McCallon}, {Tam}, {Van Dyk}, \&
  {Wheelock}}]{scs+06}
{Skrutskie}, M.~F., {Cutri}, R.~M., {Stiening}, R., {et~al.} 2006, \aj, 131,
  1163

\bibitem[{{Smith} {et~al.}(2018){Smith}, {Bruel}, {Cognard}, {Cameron},
  {Camilo}, {Dai}, {Guillemot}, {Johnson}, {Johnston}, {Keith}, {Kerr},
  {Kramer}, {Lyne}, {Manchester}, {Shannon}, {Sobey}, {Stappers}, \&
  {Weltevrede}}]{sbc+18}
{Smith}, D.~A., {Bruel}, P., {Cognard}, I., {et~al.} 2018, arXiv e-prints,
  arXiv:1812.00719

\bibitem[{{Stairs} {et~al.}(1998){Stairs}, {Arzoumanian}, {Camilo}, {Lyne},
  {Nice}, {Taylor}, {Thorsett}, \& {Wolszczan}}]{sac+98}
{Stairs}, I.~H., {Arzoumanian}, Z., {Camilo}, F., {et~al.} 1998, \apj, 505, 352

\bibitem[{{Stovall}(2013)}]{s13}
{Stovall}, K. 2013, PhD thesis, The University of Texas at San Antonio

\bibitem[{{Stovall} {et~al.}(2016){Stovall}, {Allen}, {Bogdanov}, {Brazier},
  {Camilo}, {Cardoso}, {Chatterjee}, {Cordes}, {Crawford}, {Deneva}, {Ferdman},
  {Freire}, {Hessels}, {Jenet}, {Kaplan}, {Karako-Argaman}, {Kaspi}, {Knispel},
  {Kotulla}, {Lazarus}, {Lee}, {van Leeuwen}, {Lynch}, {Lyne}, {Madsen},
  {McLaughlin}, {Patel}, {Ransom}, {Scholz}, {Siemens}, {Stairs}, {Stappers},
  {Swiggum}, {Zhu}, \& {Venkataraman}}]{sab+16}
{Stovall}, K., {Allen}, B., {Bogdanov}, S., {et~al.} 2016, \apj, 833, 192

\bibitem[{{Stovall} {et~al.}(2018){Stovall}, {Freire}, {Chatterjee},
  {Demorest}, {Lorimer}, {McLaughlin}, {Pol}, {van Leeuwen}, {Wharton},
  {Allen}, {Boyce}, {Brazier}, {Caballero}, {Camilo}, {Camuccio}, {Cordes},
  {Crawford}, {Deneva}, {Ferdman}, {Hessels}, {Jenet}, {Kaspi}, {Knispel},
  {Lazarus}, {Lynch}, {Parent}, {Patel}, {Pleunis}, {Ransom}, {Scholz},
  {Seymour}, {Siemens}, {Stairs}, {Swiggum}, \& {Zhu}}]{sfc+18}
{Stovall}, K., {Freire}, P.~C.~C., {Chatterjee}, S., {et~al.} 2018, \apjl, 854,
  L22

\bibitem[{{Strader} {et~al.}(2018){Strader}, {Swihart}, {Chomiuk}, {Bahramian},
  {Britt}, {Cheung}, {Dage}, {Halpern}, {Li}, {Mignani}, {Orosz}, {Peacock},
  {Salinas}, {Shishkovsky}, \& {Tremou}}]{ssc+18}
{Strader}, J., {Swihart}, S.~J., {Chomiuk}, L., {et~al.} 2018, arXiv e-prints,
  arXiv:1812.04626

\bibitem[{{Sugizaki} {et~al.}(2001){Sugizaki}, {Mitsuda}, {Kaneda},
  {Matsuzaki}, {Yamauchi}, \& {Koyama}}]{smk+01}
{Sugizaki}, M., {Mitsuda}, K., {Kaneda}, H., {et~al.} 2001, \apjs, 134, 77

\bibitem[{{Taam} {et~al.}(2000){Taam}, {King}, \& {Ritter}}]{tkr00}
{Taam}, R.~E., {King}, A.~R., \& {Ritter}, H. 2000, \apj, 541, 329

\bibitem[{{Tauris}(1996)}]{t96}
{Tauris}, T.~M. 1996, \aap, 315, 453

\bibitem[{{Tauris} {et~al.}(2012){Tauris}, {Langer}, \& {Kramer}}]{tlk12}
{Tauris}, T.~M., {Langer}, N., \& {Kramer}, M. 2012, \mnras, 425, 1601

\bibitem[{{Tauris} \& {Savonije}(1999)}]{ts99}
{Tauris}, T.~M., \& {Savonije}, G.~J. 1999, \aap, 350, 928

\bibitem[{{Tauris} {et~al.}(2000){Tauris}, {van den Heuvel}, \&
  {Savonije}}]{tvs00}
{Tauris}, T.~M., {van den Heuvel}, E.~P.~J., \& {Savonije}, G.~J. 2000, \apjl,
  530, L93

\bibitem[{{Tauris} {et~al.}(2017){Tauris}, {Kramer}, {Freire}, {Wex}, {Janka},
  {Langer}, {Podsiadlowski}, {Bozzo}, {Chaty}, \& {Kruckow}}]{tkf+17}
{Tauris}, T.~M., {Kramer}, M., {Freire}, P.~C.~C., {et~al.} 2017, \apj, 846,
  170

\bibitem[{{Taylor}(1992)}]{t92}
{Taylor}, J.~H. 1992, Philosophical Transactions of the Royal Society of London
  Series A, 341, 117

\bibitem[{{Taylor} {et~al.}(1979){Taylor}, {Fowler}, \& {McCulloch}}]{tfm79}
{Taylor}, J.~H., {Fowler}, L.~A., \& {McCulloch}, P.~M. 1979, \nat, 277, 437

\bibitem[{{The NANOGrav Collaboration} {et~al.}(2015){The NANOGrav
  Collaboration}, {Arzoumanian}, {Brazier}, {Burke-Spolaor}, {Chamberlin},
  {Chatterjee}, {Christy}, {Cordes}, {Cornish}, {Crowter}, {Demorest}, {Dolch},
  {Ellis}, {Ferdman}, {Fonseca}, {Garver-Daniels}, {Gonzalez}, {Jenet},
  {Jones}, {Jones}, {Kaspi}, {Koop}, {Lam}, {Lazio}, {Levin}, {Lommen},
  {Lorimer}, {Luo}, {Lynch}, {Madison}, {McLaughlin}, {McWilliams}, {Nice},
  {Palliyaguru}, {Pennucci}, {Ransom}, {Siemens}, {Stairs}, {Stinebring},
  {Stovall}, {Swiggum}, {Vallisneri}, {van Haasteren}, {Wang}, \&
  {Zhu}}]{nanograv15}
{The NANOGrav Collaboration}, {Arzoumanian}, Z., {Brazier}, A., {et~al.} 2015,
  \apj, 813, 65

\bibitem[{Valentim {et~al.}(2011)Valentim, Rangel, \& Horvath}]{vrh+11}
Valentim, R., Rangel, E., \& Horvath, J.~E. 2011, Monthly Notices of the Royal
  Astronomical Society, 414, 1427

\bibitem[{{van Leeuwen} {et~al.}(2015){van Leeuwen}, {Kasian}, {Stairs},
  {Lorimer}, {Camilo}, {Chatterjee}, {Cognard}, {Desvignes}, {Freire},
  {Janssen}, {Kramer}, {Lyne}, {Nice}, {Ransom}, {Stappers}, \&
  {Weisberg}}]{vks+15}
{van Leeuwen}, J., {Kasian}, L., {Stairs}, I.~H., {et~al.} 2015, \apj, 798, 118

\bibitem[{{Vigna-G{\'o}mez} {et~al.}(2018){Vigna-G{\'o}mez}, {Neijssel},
  {Stevenson}, {Barrett}, {Belczynski}, {Justham}, {de Mink}, {M{\"u}ller},
  {Podsiadlowski}, {Renzo}, {Sz{\'e}csi}, \& {Mandel}}]{vns+18}
{Vigna-G{\'o}mez}, A., {Neijssel}, C.~J., {Stevenson}, S., {et~al.} 2018,
  \mnras, 481, 4009

\bibitem[{{Wang} {et~al.}(2016){Wang}, {Liu}, {Qiu}, {Bai}, {Yang}, {Guo}, \&
  {Zhang}}]{wlq+16}
{Wang}, S., {Liu}, J., {Qiu}, Y., {et~al.} 2016, \apjs, 224, 40

\bibitem[{{Wolszczan}(1991)}]{w91}
{Wolszczan}, A. 1991, \nat, 350, 688

\bibitem[{{Yao} {et~al.}(2017){Yao}, {Manchester}, \& {Wang}}]{ymw16}
{Yao}, J.~M., {Manchester}, R.~N., \& {Wang}, N. 2017, \apj, 835, 29

\bibitem[{Özel {et~al.}(2012)Özel, Psaltis, Narayan, \& Villarreal}]{opn+12}
Özel, F., Psaltis, D., Narayan, R., \& Villarreal, A.~S. 2012, The
  Astrophysical Journal, 757, 55

\end{thebibliography}
